\documentclass[aps,
prb,
twocolumn,
amsmath,
amssymb,
amsfonts,
reprint,
floatfix,
superscriptaddress
]{revtex4-2}

\usepackage[utf8]{inputenc}
\usepackage[T1]{fontenc}
\usepackage[colorlinks=true, citecolor=blue, linkcolor=blue, urlcolor=blue]{hyperref}
\usepackage{chngcntr}

\usepackage{graphicx} 
\usepackage[labelformat=simple]{subcaption}
\usepackage{bm}
\usepackage{xcolor}
\usepackage{overpic}

\newcommand{\Tr}{\mathrm{Tr}}

\begin{document}
\title{Relevant long-range interaction of the entanglement Hamiltonian emerges from a short-range gapped system}
\author{Chuhao Li}
\affiliation{Beijing National Laboratory for Condensed Matter Physics and Institute of Physics, Chinese Academy of Sciences, Beijing 100190, China}
\affiliation{School of Physical Sciences, University of Chinese Academy of Sciences, Beijing 100190, China}

\author{Rui-Zhen Huang}
\affiliation{Department of Physics and Astronomy, Ghent University, Krijgslaan 281, S9, B-9000 Ghent, Belgium}

\author{Yi-Ming Ding}
\affiliation{State Key Laboratory of Surface Physics and Department of Physics, Fudan University, Shanghai 200438, China}
\affiliation{Department of Physics, School of Science and Research Center for Industries of the Future, Westlake University, Hangzhou 310030,  China}
\affiliation{Institute of Natural Sciences, Westlake Institute for Advanced Study, Hangzhou 310024, China}

\author{Zi Yang Meng}
\affiliation{Department of Physics and HKU-UCAS Joint Institute of Theoretical and Computational Physics,The University of Hong Kong, Pokfulam Road, Hong Kong SAR, China}

\author{Yan-Cheng Wang}
\email{ycwangphys@buaa.edu.cn}
\affiliation{Hangzhou International Innovation Institute, Beihang University, Hangzhou 311115, China}
\affiliation{Tianmushan Laboratory, Hangzhou 311115, China}

\author{Zheng Yan}
\email{zhengyan@westlake.edu.cn}
\affiliation{Department of Physics, School of Science and Research Center for Industries of the Future, Westlake University, Hangzhou 310030,  China}
\affiliation{Institute of Natural Sciences, Westlake Institute for Advanced Study, Hangzhou 310024, China}

\begin{abstract}
Beyond the Li-Haldane-Poilblanc conjecture, we find the entanglement Hamiltonian (EH) is actually not closely similar to the original Hamiltonian on the virtual edge. Unexpectedly, the EH has some relevant long-range interacting terms which hugely affect the physics. Without loss of generality, we study a spin-1/2 Heisenberg bilayer to obtain the entanglement information between the two layers through our newly developed quantum Monte Carlo scheme, which can simulate large-scale EH. Although the entanglement spectrum carrying the Goldstone mode seems like a Heisenberg model on a single layer, which is consistent with Li-Haldane-Poilblanc conjecture, we demonstrate that there actually exists a finite-temperature phase transition of the EH. The results violate the Mermin-Wagner theorem, which means there should be relevant long-range terms in the EH. It reveals that the Li-Haldane-Poilblanc conjecture ignores necessary corrections for the EH which may lead totally different physics.
\end{abstract}

\maketitle

\section{Introduction}
Quantum entanglement is a powerful tool to extract and characterize the informational, field-theoretical, and topological properties of quantum many-body systems~\cite{vidal2003entanglement,Korepin2004universality,Kitaev2006,Levin2006}, which combines the conformal field theory (CFT) and the categorical description of the problem~\cite{Calabrese2008entangle,Fradkin2006entangle,Nussinov2006,Nussinov2009,CASINI2007,JiPRR2019,ji2019categorical,kong2020algebraic,XCWu2020,XCWu2021,JRZhao2021,JRZhao2022,10.21468/SciPostPhys.13.6.123}.
Low-lying entanglement spectrum (ES) has been widely employed as a fingerprint of CFT and topology in the investigation in highly entangled quantum matter~\cite{Pollmann2010entangle,Fidkowski2010,Yao2010,XLQi2012,Canovi2014,LuitzPRL2014,LuitzPRB2014,LuitzIOP2014,Chung2014,Pichler2016,Cirac2011,Stoj2020,guo2021entanglement,Grover2013,Assaad2014,Assaad2015,Parisen2018,PhysRevLett.129.210601,Moradi_2016}. 

Besides the famous gapped phase in the spin-integer antiferromagnetic Heisenberg chain as Haldane's conjecture~\cite{haldaneContinuum1983,haldaneNonlinear1983}, there is another well-noted Haldane's conjecture about the relationship between the entanglement spectrum and edge energy spectrum. More than one decade ago, Li and Haldane creatively pointed out the ES may be a more precise physical quantity rather than entanglement entropy (EE)~\cite{Li2008entangle}. 
Furthermore, they demonstrated that the general $\nu=5/2$ topological states have the same low-lying ES to identify the topology and CFT structure. In addition, they predicted that the ES of the topological state would be very similar to the energy spectrum of the edge state, that is, the Li-Haldane conjecture (Haldane's conjecture for ES). The ES is defined as the energy spectrum of the corresponding entanglement Hamiltonian (EH). If a total system is separated into subsystem $A$ and environment $\bar{A}$, then the EH of $A$, i.e., $H_A$, can be written as $H_A=-\ln\rho_A$, where the $\rho_A=\Tr_{\bar{A}}(\rho)$ is the reduced density matrix (RDM) of the subsystem $A$.

The conjecture seems not only limited to the topological states. Two years later, Poilblanc pointed out, via numerical results, that the relation between the low-lying ES and edge energy spectrum exists generally in quantum spin systems beyond topological states~\cite{Poilblanc2010entanglement}. Thereafter, the conjecture has also been called as Li-Haldane-Poilblanc conjecture and extended into general cases beyond topological systems. Then, Qi, Katsura and Ludwig theoretically proved the general relationship between ES of $(2+1)$d gapped topological states and the spectrum on their $(1+1)$d edges exactly when the edge is a CFT~\cite{XLQi2012}. 
\begin{figure*}
    \centering
    \begin{subcaptiongroup}
    	\begin{overpic}[width=0.98\textwidth]{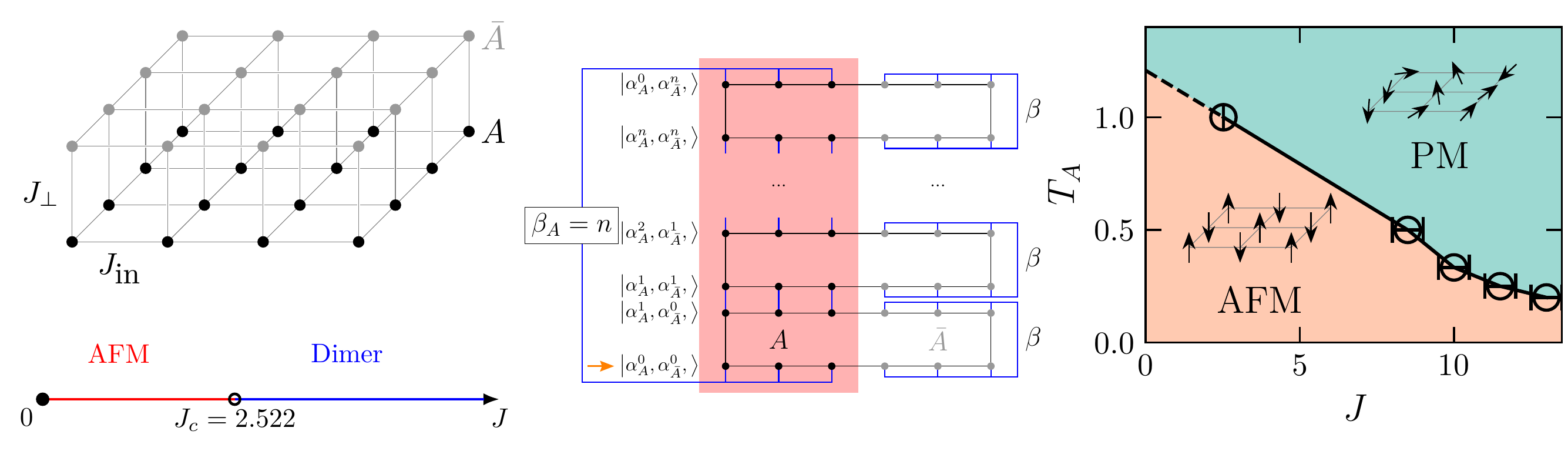}
    		\subcaptionlistentry{Bilayer_Heisenberg_Lattice}
    		\put(0,24.5){\label{fig:Bilayer_Heisenberg_Lattice} {\Large \captiontext*{}}}
    		\subcaptionlistentry{replica_geometry}
    		\put(35,24.5){\label{fig:replica_geometry} {\Large \captiontext*{}}}
    		\subcaptionlistentry{n-J plot}
    		\put(73,24.5){\label{fig:n_Jc} {\Large \captiontext*{}}}
    	\end{overpic}
    \end{subcaptiongroup}
    \caption{
    \subref{fig:Bilayer_Heisenberg_Lattice} The lattice of the bilayer Heisenberg model. Darker dots denote sites in the region $A$ we are concerned with. Lighter dots denote sites in the environment $\bar{A}$. 
    The lower part shows the quantum phase diagram of the vanilla bilayer Heisenberg model, which has a quantum phase transition from the AFM phase to the dimer phase, and its critical point is located at $J_c=2.522$; the result comes from Refs.~\cite{Wang2006bilayer, Lohoefer2015}.
    \subref{fig:replica_geometry} An illustration of the replica manifold structure in the imaginary time of the QMC simulation. The blue line indicates the connected spin states which are identical on the edge of the replicas. The spin states in region $A$ and $\bar{A}$ are connected differently in their own region, and follow a different PBC. The orange arrow indicates the imaginary time to do the MC measurement.
    \subref{fig:n_Jc} The phase diagram of the entanglement Hamiltonian $H_A$. The $J=J_{\bot}/J_{\textrm{in}}$ is the coupling ratio of the original Hamiltonian of the bilayer Heisenberg model. The $T_{\textrm{A}}$ is the effective temperature of the entanglement Hamiltonian $H_A$, that is, the partition function of the entanglement Hamiltonian is $Z=e^{-H_A/T_{\textrm{A}}}.$
    The data of the points in the plot are listed in Table.\ref{tab:Crtical_J}. The point of $T_A = 1$ corresponds to the $J_c$ in the lower parts of \subref{fig:Bilayer_Heisenberg_Lattice}.
    The dashed line in \subref{fig:n_Jc} is directly extended to $J=0$ from the segment established by the points of $T_A = 1$ and $T_A = 0.5$.
    }
    \label{fig:1}
\end{figure*}

Recent studies have explored the EH in greater depth, using a combination of field theory, numerical simulations, and experimental data.~\cite{alba2012entanglement,PhysRevLett.110.260403,zhu2020entanglement,zhu2019reconstructing,Tang2020critical,Mendes_Santos_2020,Dalmonte_2022,Joshi2023,redon2023realizing,dalmonte2018quantum,ma2023multipartite,Eisler_2020,PhysRevB.100.195109,Cardy_2016,Javerzat_2022,Eisler_2019}. The correspondence between edge and entanglement spectra has been successfully applied to many quantum states of matter with topological properties \cite{Critical,Entanglement2011,Chiral,classification}. 
Previous studies on a one-dimensional (1D) system implies the EH might not strictly be a short-range one, but instead involves long-range interactions characterized by exponential decay. Remarkably, these long-range interactions appear to have little impact on the short-range entanglement dynamics. The low-lying ES in these systems shows similarities with both the EH and the edge energy spectrum, leading us to consider that the long-range terms of the entanglement Hamiltonian might be insignificant and can potentially be ignored for an approximate analysis~\cite{lauchli2012entanglement,zhu2019reconstructing}.

Given that the EH and the edge Hamiltonian have been considered comparable in many contexts, it is reasonable to expect that they share fundamental physical properties. At a minimum, their basic features should exhibit similarities.
However, whether the seemingly useless long-range terms are indeed irrelevant is still ignored under the conventional wisdom. 

Meanwhile, the influence of long-range interactions has been widely studied in past decades~\cite{PhysRevResearch.5.033038,PhysRevResearch.5.033046,chiocchetta2021cavity,Yusuf2004spin,defenu2021metastability,Birnkammer2022Characterizing,Horita2017upper,FUKUI20092629}. It is well known that the long-range interactions $\sim \frac 1 {r^\alpha}$ are irrelevant when the decaying power is large. Otherwise, the long-range terms indeed change the intrinsic physics, e.g., violating the Mermin-Wagner theorem~\cite{mermin1966absence,mermin1967absence} even then destroying the Goldstone mode in continuous symmetry breaking systems~\cite{Anderson1952,Oguchi1960,Shao2017nearly,Zhou2021amplitude,PhysRevB.105.014418}. Thus the Mermin-Wagner theorem is a good standard to check whether the long-range interaction is relevant. The difficulty lies in how to extract the information of the EH in a $(2+1)$d entangled system. It was extremely challenging for previous numerical methods, such as exact diagonalization (ED) and density matrix renormalization group (DMRG) algorithms due to the limited system sizes. We note the authors of Ref.\cite{Chandran2014how} pointed out the existing thermal phase transition for EH to question the universality of the ES. However, the discussion in the reference is still located in the region of Mermin-Wagner theorem and defaults to the EH are short ranged.

Recently, two of the authors proposed a numerical scheme by designing an $n$th R\'enyi RDM $\rho_A^n=e^{-n\mathcal{H}_{A}}$ within the path-integral frame to capture the information of the entanglement spectrum $\mathcal{H}_{A}$ by treating the large $n$ as an effective imaginary time length $\beta_A$ (reciprocal of temperature). A replica partition function of $n$th R\'enyi RDM $\mathcal{Z}^{(n)}_{A}=\Tr[\rho^{n}_{A}]=\Tr[e^{-n\mathcal{H}_{A}}]$ was constructed and can be simulated via quantum Monte Carlo (QMC)~\cite{zyan2021entanglement,Assaad2014,Assaad2015,Parisen2018}.
More details to calculate the ES from the QMC can be found in Ref.~\cite{zyan2021entanglement}. In that method, the $n$ is fixed in a large value to simulate the imaginary time evolution of the EH. Inspired by it, we find the finite-temperature information of EH can be obtained by the QMC if we set the $n$ as a variable. Thus the problem proposed above can now be studied.

\begin{figure*}[htb]
    \centering
    \begin{subcaptiongroup}
        \begin{overpic}[width=0.99\textwidth]{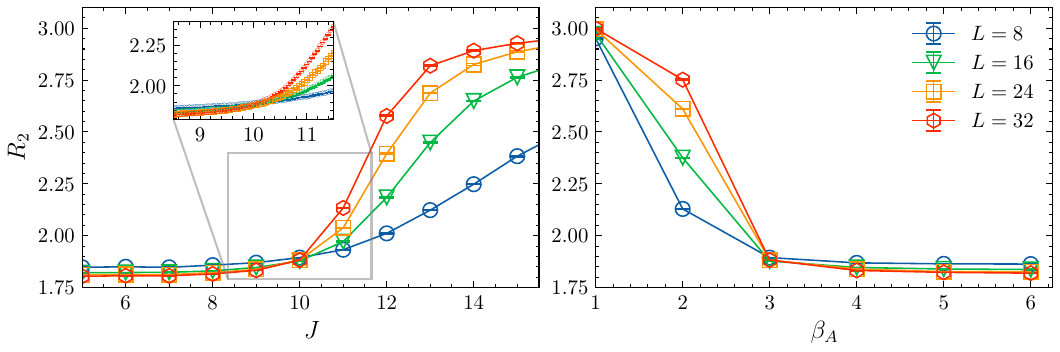}
            \subcaptionlistentry{R2_n3_comp_l_linear}
            \put(10,12){\label{fig:R2_n3_comp_l_linear} {\Large \captiontext*{}}}
            \subcaptionlistentry{R2_J10_comp_l_linear}
            \put(57,12){\label{fig:R2_J10_comp_l_linear} {\Large \captiontext*{}}}
        \end{overpic}
    \end{subcaptiongroup}

    \caption{
    The Binder ratio $R_2$ results of different $L$ \subref{fig:R2_n3_comp_l_linear} under $\beta_A=3$ sweeping $J$ and \subref{fig:R2_J10_comp_l_linear} under $J=10.0$ sweeping $\beta_A$.
    The crossing point in \subref{fig:R2_n3_comp_l_linear} indicates a phase transition nearing $J=10.0$. The inset in \subref{fig:R2_n3_comp_l_linear} shows more detailed data around the crossing point, ranging from $J=8.5$ to $J=11.5$ and spacing with $0.05$.
    While using $J=10.0$ in \subref{fig:R2_J10_comp_l_linear}, the crossing point locates at $\beta_A=3$ approximately, which also proves the existence of the phase transition.
    All calculations are under low real temperature $\beta=2 \times L$.
    }
    \label{fig:2}
\end{figure*}

\section{Model}
To demonstrate the relevant difference between the edge Hamiltonian and the EH, we take a $S=1/2$ antiferromagnetic Heisenberg model on a bilayer square lattice as an example. The Hamiltonian can be written as
\begin{equation}
    \label{eqn:Hamiltonian}
    H = J_{\textrm{in}} \sum_{\langle ij \rangle} (\bm{S}_{i,1} \bm{S}_{j,1} + \bm{S}_{i,2} \bm{S}_{j,2}) + J_{\bot} \sum_i \bm{S}_{i,1} \bm{S}_{i,2} ,
\end{equation}
where $J_{\textrm{in}}$ denotes the intralayer interaction constant and $J_{\bot}$ denotes the interlayer interaction constant. Here the $\bm{S}_{i,l}$ is the spin-1/2 operator at site $i$ and layer $l\ (l=1,2)$. Each layer is a periodic boundary condition (PBC) square lattice with sites of $N=L\times L$, and $\langle i j \rangle$ represents the intralayer pair of nearest-neighbor sites. We next define $J=J_{\bot}/J_{\textrm{in}}$ to reflect the relative strength of the interaction.

The ground-state phase diagram of such a bilayer model has been well studied in previous works: the $(2+1)$d $O(3)$ quantum critical point(QCP), separating the N\'eel phase and inter-layer dimerized phase, is found to be located at $J=2.5220(1)$ from high-precision QMC simulations~\cite{Wang2006bilayer,Lohoefer2015,JRZhao2021,10.21468/SciPostPhys.13.6.123}, as shown in the lower part of Fig.~\ref{fig:Bilayer_Heisenberg_Lattice}. 
We choose one layer to be the subsystem $A$ and the other one to be the environment $\bar{A}$. Previous studies hinted that the entanglement Hamiltonian $\mathcal{H}_A$ would resemble a Heisenberg model on the square lattice following the Li-Haldane-Poilblanc conjecture~\cite{Li2008entangle,Poilblanc2010entanglement}. 

Unexpectedly, we find that there exists a continuous phase transition in the EH at the finite effective temperature $T_A$. 
The system undergoes a spontaneous symmetry breaking from high to low temperature. The phase diagram is shown in Fig.~\ref{fig:n_Jc}. Our finding suggests that the EH can not simply be understood as a single-layer Heisenberg model, in which the Mermin-Wagner theorem forbids the spontaneous breaking of continuous symmetries at a finite temperature in a short-range interacting model. However, such a continuous phase transition can happen when the interactions become long range~\cite{Zhao2023,song2023quantum,PhysRevResearch.5.033046}. Therefore, our results strongly indicate that some important corrections between the EH and the edge Hamiltonian are needed necessarily. 

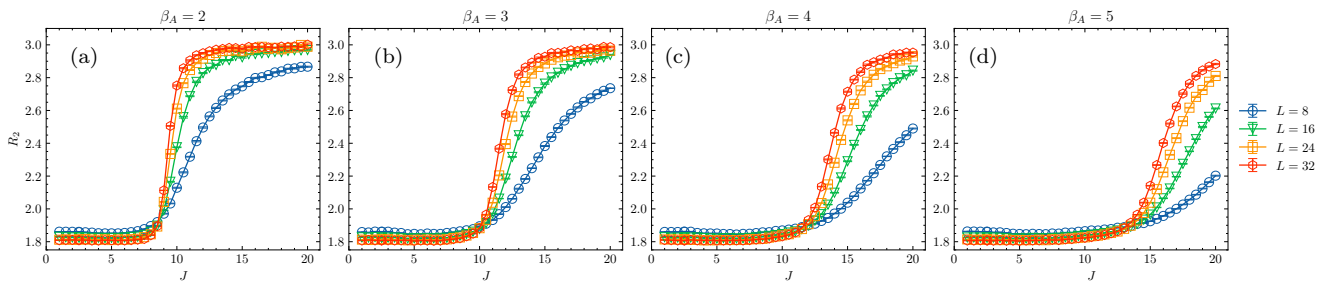
\begin{figure*}
    \centering
    \begin{subcaptiongroup}
    \begin{overpic}[width=0.98\textwidth]{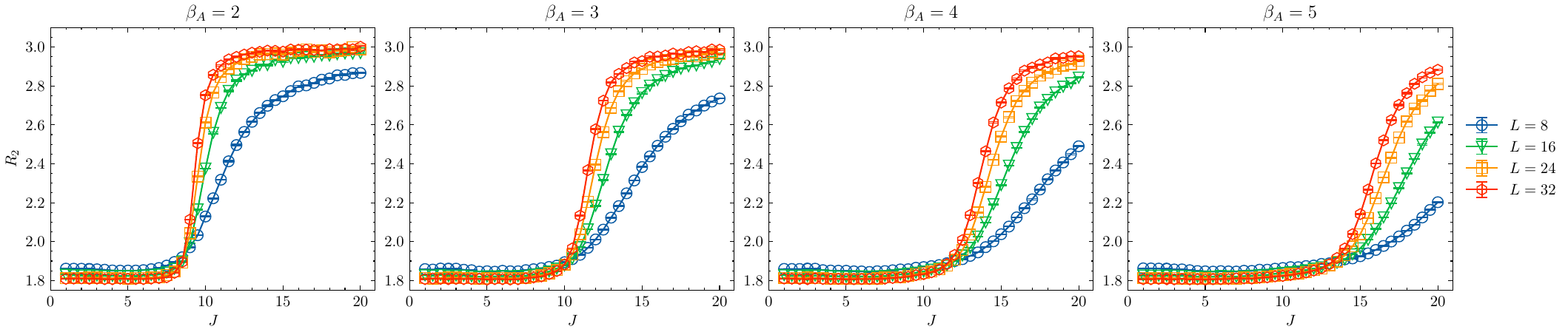}
        \subcaptionlistentry{U2_n2}
        \put (5,17) {\captiontext*{}}
        \label{fig:U2_n2}
        \subcaptionlistentry{U2_n3}
        \put (28,17) {\captiontext*{}}
        \label{fig:U2_n3}
        \subcaptionlistentry{U2_n4}
        \put (50,17) {\captiontext*{}}
        \label{fig:U2_n4}
        \subcaptionlistentry{U2_n5}
        \put (73,17) {\captiontext*{}}
        \label{fig:U2_n5}
    \end{overpic}
    \end{subcaptiongroup}
    \caption{The Binder ratio $R_2$ results at \subref{fig:U2_n2} $\beta_A = 2$, \subref{fig:U2_n3} $\beta_A = 3$, \subref{fig:U2_n4} $\beta_A = 4$, \subref{fig:U2_n5} $\beta_A = 5$, with a wide range sweeping of $J$ from $J=1$ to $J=20$, and comparing the system size of $L=8,16,24,32$. These results suggest that a finite temperature phase transition exists on EH.}
    \label{fig:fig3}
\end{figure*}

\section{Method}
The central problem is how to simulate the $n$th R\'enyi RDM $\Tr[e^{-n\mathcal{H}_{A}}]$ by QMC. As previous research~\cite{zyan2021entanglement} suggests, $\mathcal{Z}_{A}^{(n)}$ is a partition function in a replicated manifold, where the time boundaries of region $A$ of the $n$ replicas are connected along imaginary time and the time boundaries of the region $\overline{A}$ of every replica are independent to each other (for sites in $\overline{A}$ for each replica, the usual periodic boundary condition of $\beta$ is maintained). 

{
We then provide a brief overview of how we applied the simulation method proposed in Ref.~\cite{zyan2021entanglement} to our study.
We start by writing the state of the system as $\left| \alpha_{A}^{i}, \alpha_{\bar{A}}^{j} \right\rangle$, in which we are using the superscripts $i,j=0,1,\cdots,n\ (n=\beta_A)$ to denote the index of the replica and subscripts $A$ and $\bar{A}$ to denote the region. With the trace relationship mentioned above and swapping the order of trace, we now have $ Z^{(n)}_A = \Tr[\rho_A^n]= \Tr[(\Tr_{\bar{A}} \rho)^n] = \Tr_{\bar{A}}[(\Tr\rho)^n]$. After expanding the inner trace, we could get
\begin{equation}
    \begin{aligned}
        (\Tr\rho)^n &= \langle \alpha^0_A, \alpha^0_{\bar{A}} | e^{-\beta H} | \alpha^1_A, \alpha^0_{\bar{A}} \rangle \\
        &\langle \alpha^1_A, \alpha^1_{\bar{A}} | e^{-\beta H}|
        \alpha^2_A, \alpha^1_{\bar{A}} \rangle \\
 &\langle \alpha^2_A, \alpha^2_{\bar{A}} | e^{-\beta H}|
        \alpha^3_A, \alpha^2_{\bar{A}} \rangle \\
        & \cdots \\
        &  \langle \alpha^i_A, \alpha^i_{\bar{A}} | e^{-\beta H}|
        \alpha^i_A, \alpha^{(i-1)}_{\bar{A}} \rangle\\
        & \cdots \\
         &\langle \alpha^n_A, \alpha^n_{\bar{A}} |e^{-\beta H}| \alpha^0_{A},\alpha^n_{\bar{A}} \rangle.
    \end{aligned}
\end{equation}
Noticing that we already utilized the relationship that, at the edge of two adjacent $e^{-\beta H}$ in the imaginary time of region $A$, $\left|\alpha_{A}^i \right\rangle = \left|\alpha_{A}^{(i-1)} \right\rangle$, and the PBC of region $A$ is maintained as well.} The whole process is also depicted as Fig.~\ref{fig:replica_geometry}.

Therefore, the finite temperature properties of the EH can be extracted numerically. Here $n$ performs the role of an effective inverse temperature $\beta_A$ for the EH. In this way, we can readily make use of such effective imaginary time $\beta_A=n$ at those $n=1,2,3,\dots$, integer points to extract the thermal properties of the EH. 

It's important to highlight that the term ``temperature'' can refer to two distinct concepts in this context. The first is the actual temperature $T$, related to the original Hamiltonian $H$, with $T = 1/\beta$, and the other one is the effective temperature $T_A$ associated with the EH $\mathcal{H}_{A}$, where $T_A = 1/\beta_A$. It can be seen that the effective inverse temperature $\beta_A= n$ for the EH $\mathcal{H}_{A}$ of the subsystem $A$ is in the unit of $1$ whereas the $\beta=1/T$ of the total system is in the inverse unit of the physical energy scale of the original system, $J$ of the Heisenberg model, for instance. Here we set the $\beta$ to be large and allow it to increase proportionally with the system size $L$ to make the system approach its ground state. When discussing the finite temperature phase transition of the EH $\mathcal{H}_{A}$, the ``temperature'' here refers to the $T_A$ ($1/\beta_A$). In the replica manifold we simulate, the $\beta_A$ can be tuned by changing the number of replicas $n$. 

As mentioned above, with the expansion form of partition function $\mathcal{Z}_A^{(n)}$, we could then construct a special geometry structure in the stochastic series expansion(SSE) QMC~\cite{sandvik1998stochastic,sandvik2010computational,Sandvik1991,Sandvik1999,Syljuaasen2002,ZY2019,yan2020improved} to measure those order parameters. All the measurements here are defined in the scope of EH. As an example, for a physical observable $O$ defined on EH and its corresponding partition function $Z^{(n)}_A$, the expectation value $\langle O \rangle_{H_{A}}$ is given by
\begin{equation}
    \langle O \rangle_{H_{A}} = \Tr[\rho_A^n O].
\end{equation}
Applying the relationship
\begin{equation}
    e^{-H_{A}} = \rho_A = \Tr_{\bar{A}} \rho = \Tr_{\bar{A}} [e^{-\beta H}],
\end{equation}
one can write it as 
\begin{equation}
    \langle O \rangle_{H_{A}} = \Tr[ (\Tr_{\bar{A}} [e^{-\beta H}])^n O ].
\end{equation}
Note that according to the property of trace calculation, all the measurements about the EH should be done in the connection of the two neighboring  replicas~\cite{zyan2021entanglement,liu2023probing,wu2023classical,song2023different}. For simplicity, the physical observables in this work are taken on the $|\alpha^0_{A}\rangle$ as the arrow points out in Fig.\ref{fig:replica_geometry}.

\section{Results}
Since we established the basic method, we then perform the simulation with 12 independent processes. Each process executes 5000 MC steps of warming up and the measurement value is averaged over 20 bins, with 10000 MC steps in each. The result is then averaged by those 12 results.

\subsection{Finite-temperature phase transition of the EH}

Firstl, we measure the order parameter $\langle m^2\rangle = \langle(\frac{1}{N} \sum_{i=1}^{N} \phi_i S^{z}_{i,1})^2\rangle$,
where $\phi_i = (-1)^{x_i+y_i}$. 
Then from the magnetization operator, we can also calculate the corresponding second Binder ratio $R_2$ for the order parameter of the EH, which is defined as $R_2 = \langle m^4 \rangle/ \langle m^2\rangle^2$.

Considering our method is limited to the discrete $\beta_A$, we next fix the $\beta_A$ and scan the continuous $J$ to obtain better precision results. Therefore, the finite-temperature phase transition points at fixed $\beta_A$ can be calculated much more accurately. 
The results of the Binder ratio $R_2$ under $\beta_A=3$ are shown in Fig.~\ref{fig:R2_n3_comp_l_linear} as an example, within the length of system size $L=8, 16, 24, 32$. From Fig.~\ref{fig:R2_n3_comp_l_linear}, we can locate the critical point at $J=10.0(5)$ under $\beta_A=3$. Then a calculation sweeping integer $\beta_A$ ($n$) is then carried with same system sizes configurations, to show that those curves indeed cross at $\beta_A=3$, as an additional proof. The results are plotted in Fig.~\ref{fig:R2_J10_comp_l_linear}.

\begin{figure*}[tb]
    \centering
    \begin{subcaptiongroup}
        \begin{overpic}[width=0.99\textwidth]{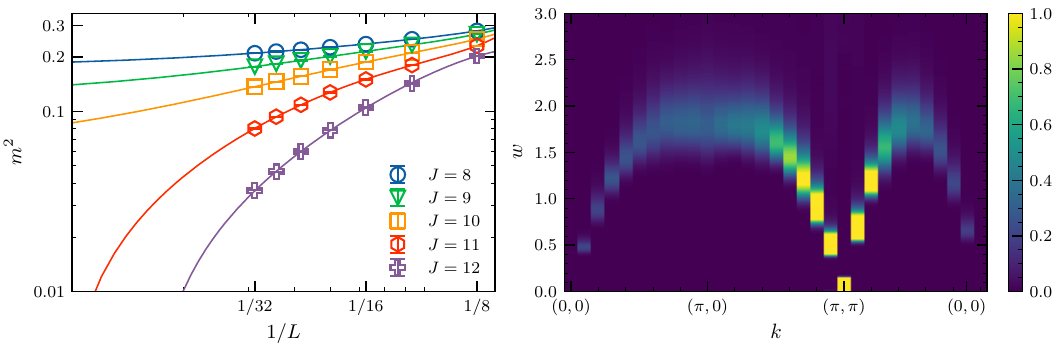}
            \subcaptionlistentry{Fig4-a}
            \put(9,29){\label{fig:n3_extrapolate} {\Large \captiontext*{}}}
            \subcaptionlistentry{Fig4-b}
            \put(54,29){\label{fig:Sw_L20_b40_J10} {\color{white} \Large \captiontext*{}}}
        \end{overpic}
    \end{subcaptiongroup}
    \caption{\subref{fig:n3_extrapolate} Finite-size scaling of $m^2$ under $\beta_A=3$, with $J = 8, 9,10,11,12$. All curves are fitted with a third order polynomial, and then extrapolated to nearing 0.
    The extrapolate shows a highly possible signal that there is a phase transition between $J=10$ and $J=11$.
    \subref{fig:Sw_L20_b40_J10} Entanglement spectrum $S(\mathbf{k},\omega)$ of a replicated $L=20$ bilayer antiferromagnetic Heisenberg model with $\beta=2\times L$, $\beta_A=50$ and $J=10$. The maximum values of $S(\mathbf{k},\omega)$ are truncated to 1.}
    \label{fig:Fig4}
\end{figure*}

The finite-temperature phase transition point of the EH at $\beta_A=1$ is the same as the QCP of the original Hamiltonian at zero temperature. They share the same $J_c \sim 2.52$. 
From the quantum phase transition view, this criticality is a $(2+1)$d $O(3)$ class. It implies that the corresponding two-dimensional (2D) thermal dynamical phase transition of the EH should involve others (relevant long-range terms) to make it seems like a $(2+1)$d criticality.

We also expand the calculation of the Binder ratio $R_2$ to other $\beta_A$s, those results at $\beta_A=2,3,4,5$ are shown as Fig.~\ref{fig:fig3}. Then we could locate those $\beta_A$'s corresponding critical $J$s and arrange them as the table in Table.~\ref{tab:Crtical_J}. These data points form the phase diagram of the EH shown in Fig.~\ref{fig:n_Jc}.

\begin{table}[htb]
    \begin{tabular}{cccccc}
        \hline
        $\beta_A$ & 1 & 2 & 3 & 4 & 5 \\
        $J$ & 2.5220(1)  &  8.5(5) & 10.0(5)  &  11.5(5) &  13.0(5) \\ 
        \hline
    \end{tabular}
    \caption{The critical $J$s under different $\beta_A$. The high precision result of critical $J$ under $\beta_A=1$ comes directly from Ref.\cite{Wang2006bilayer}, as a reference.}
    \label{tab:Crtical_J}
\end{table}

\subsection{Stable order of the EH at low temperature}
Due to the limit of the integer $\beta_A$, the finite temperature phase transition may not be solid enough via the above evidence in the readers' opinions. Hereby, through fixing the $\beta_A=3$, we carefully calculate the order parameter under different coupling ratio $J$ and system sizes $L$, as shown in Fig.~\ref{fig:n3_extrapolate} with double logarithmic axes. Here we also set the original reciprocal-temperature $\beta$ scales with system size $L$ where $\beta = 2\times L$, to avoid the finite size/temperature effect.

It is obvious that the $\langle m^2 \rangle$ remains a finite value of $J=8$ and $9$. The data of $J=10$ resemble a straight line under double logarithmic axes which is also another evidence for the critical point consistent to the Fig.~\ref{fig:2}. 
These results are also consistent with the phase diagram of the EH in Fig.~\ref{fig:n_Jc}: while the coupling ratio $J$ becomes smaller, the critical temperature $T_A$ rises higher. The ordered phase at low temperature of the EH is confirmed without any doubts, which also reflects the convinced finite temperature phase transition of the EH. According to the symmetry analysis of the system, the EH must be constructed via SU(2) terms to keep its continuous symmetry. All the evidence now support this 2D phase transition of continuous symmetry breaking beyond the Mermin-Wagner theorem, which reveals the exist of relevant long-range terms.

\subsection{Entanglement spectrum}
The EH actually exhibits interactions that significantly deviate from those predicted by the well-known Li-Haldane-Poilblanc conjecture. The EH has a finite temperature phase transition while the edge Hamiltonian has nothing. The remaining question becomes the following: How large is the difference between the entanglement spectrum and the edge energy spectrum in this case.
Combined with the QMC calculation of the RDM and stochastic analytic continuation (SAC)~\cite{Sandvik1998a,Beach2004,Syljuasen2008,SHAO2023Progress}, we can obtain the entanglement spectrum using the scheme of Refs.\cite{zyan2021entanglement,liu2023probing,wu2023classical,song2023different}. As shown in Fig.~\ref{fig:Sw_L20_b40_J10}, the entanglement spectrum seems still similar to a spin-wave excitation of a short-range Heisenberg model on a square lattice, i.e., the edge Hamiltonian. From this aspect, the entanglement spectrum approximately resembles the edge Hamiltonian at least visually. That is, the relevant long-range interaction obviously changes the property of the physical system to induce a finite temperature phase transition, but it still has not hugely modified the structure of the spectra information.
It may be the main reason why such a difference between the EH and edge Hamiltonian has not been found in a long time.

\section{Conclusion and discussion}
In this work, we find that, although the spectrum structures of the entanglement Hamiltonian and the edge Hamiltonian are almost similar, which obeys the Li-Haldane-Poilblanc conjecture, the actual properties of the two Hamiltonians are totally different. 
Taking the Heisenberg model on bilayer square lattice as an example and setting one layer as environment, the 2d entanglement Hamiltonian containing continuous symmetry exhibits a finite temperature phase transition which violates the Mermin-Wagner theorem. Via designing the path integral manifold of QMC, we have simulated the finite temperature property of the EH to strongly support above conclusion.
All the evidence points to the fact that there should be some necessary corrections with long-range interacting terms to the EH. From the spectra analysis, although the corrections 
hugely change the physical properties of the EH and make it totally different from the edge Hamiltonian, the two Hamiltonians' spectra are still seemly semblable.

We note that a gapless ground state with long-range entanglement may lead to a long-range EH in some analytic arguments~\cite{PhysRevA.103.043321,PhysRevB.88.245137}. A very recent theoretical work (almost the same time as ours) concluded a trivial gapped state (e.g., dimer phase) can not hold a long range EH \cite{zhou2023reviving} and the conclusion was demonstrated numerically in 1D systems \cite{zhou2023reviving,li2023numerical}. In fact, our result provides an interesting counterexample in two dimensions, thus attracting more further research.

\begin{acknowledgments}
We acknowledge useful discussions with Zheng Zhou, Bin-Bin Chen, Meng Cheng, Xiao-Liang Qi and Wei Zhu. We also appreciate the help from Weilun Jiang and Gaopei Pan.
YMD and ZY acknowledge the support from the start-up funding of Westlake University.
R.Z.H is supported by a postdoctoral fellowship from the Special Research Fund (BOF) of Ghent University.
ZYM thanks support from the RGC of Hong Kong SAR of China (Projects No. 17301420, No. 17301721, No. AoE/P-701/20, No. 17309822, and No. HKU C7037-22G), the ANR/RGC Joint Research Scheme sponsored by Research Grants Council of Hong Kong SAR of China and French National Research Agency(Project No. A\_HKU703/22), the K. C. Wong Education Foundation (Grant No.~GJTD-2020-01), and the Seed Funding ``Quantum-Inspired explainable-AI'' at
the HKU-TCL Joint Research Centre for Artificial Intelligence. 
Y.C.W. acknowledges support from the Zhejiang Provincial Natural Science Foundation of China (Grant No. LZ23A040003). 
We thank Beijng PARATERA Tech Co., Ltd., 
the Blackbody HPC system at the Department of Physics, University of Hong Kong, HPC centre of Westlake University for providing computational resources that  contributed to the research results in this paper.
\end{acknowledgments}

\bibliography{ES}

\begin{thebibliography}{100}%
\makeatletter
\providecommand \@ifxundefined [1]{%
 \@ifx{#1\undefined}
}%
\providecommand \@ifnum [1]{%
 \ifnum #1\expandafter \@firstoftwo
 \else \expandafter \@secondoftwo
 \fi
}%
\providecommand \@ifx [1]{%
 \ifx #1\expandafter \@firstoftwo
 \else \expandafter \@secondoftwo
 \fi
}%
\providecommand \natexlab [1]{#1}%
\providecommand \enquote  [1]{``#1''}%
\providecommand \bibnamefont  [1]{#1}%
\providecommand \bibfnamefont [1]{#1}%
\providecommand \citenamefont [1]{#1}%
\providecommand \href@noop [0]{\@secondoftwo}%
\providecommand \href [0]{\begingroup \@sanitize@url \@href}%
\providecommand \@href[1]{\@@startlink{#1}\@@href}%
\providecommand \@@href[1]{\endgroup#1\@@endlink}%
\providecommand \@sanitize@url [0]{\catcode `\\12\catcode `\$12\catcode
  `\&12\catcode `\#12\catcode `\^12\catcode `\_12\catcode `\%12\relax}%
\providecommand \@@startlink[1]{}%
\providecommand \@@endlink[0]{}%
\providecommand \url  [0]{\begingroup\@sanitize@url \@url }%
\providecommand \@url [1]{\endgroup\@href {#1}{\urlprefix }}%
\providecommand \urlprefix  [0]{URL }%
\providecommand \Eprint [0]{\href }%
\providecommand \doibase [0]{https://doi.org/}%
\providecommand \selectlanguage [0]{\@gobble}%
\providecommand \bibinfo  [0]{\@secondoftwo}%
\providecommand \bibfield  [0]{\@secondoftwo}%
\providecommand \translation [1]{[#1]}%
\providecommand \BibitemOpen [0]{}%
\providecommand \bibitemStop [0]{}%
\providecommand \bibitemNoStop [0]{.\EOS\space}%
\providecommand \EOS [0]{\spacefactor3000\relax}%
\providecommand \BibitemShut  [1]{\csname bibitem#1\endcsname}%
\let\auto@bib@innerbib\@empty
\bibitem [{\citenamefont {Vidal}\ \emph {et~al.}(2003)\citenamefont {Vidal},
  \citenamefont {Latorre}, \citenamefont {Rico},\ and\ \citenamefont
  {Kitaev}}]{vidal2003entanglement}%
  \BibitemOpen
  \bibfield  {author} {\bibinfo {author} {\bibfnamefont {G.}~\bibnamefont
  {Vidal}}, \bibinfo {author} {\bibfnamefont {J.~I.}\ \bibnamefont {Latorre}},
  \bibinfo {author} {\bibfnamefont {E.}~\bibnamefont {Rico}},\ and\ \bibinfo
  {author} {\bibfnamefont {A.}~\bibnamefont {Kitaev}},\ }\bibfield  {title}
  {\bibinfo {title} {Entanglement in quantum critical phenomena},\ }\href
  {https://doi.org/10.1103/PhysRevLett.90.227902} {\bibfield  {journal}
  {\bibinfo  {journal} {Phys. Rev. Lett.}\ }\textbf {\bibinfo {volume} {90}},\
  \bibinfo {pages} {227902} (\bibinfo {year} {2003})}\BibitemShut {NoStop}%
\bibitem [{\citenamefont {Korepin}(2004)}]{Korepin2004universality}%
  \BibitemOpen
  \bibfield  {author} {\bibinfo {author} {\bibfnamefont {V.~E.}\ \bibnamefont
  {Korepin}},\ }\bibfield  {title} {\bibinfo {title} {Universality of entropy
  scaling in one dimensional gapless models},\ }\href
  {https://doi.org/10.1103/PhysRevLett.92.096402} {\bibfield  {journal}
  {\bibinfo  {journal} {Phys. Rev. Lett.}\ }\textbf {\bibinfo {volume} {92}},\
  \bibinfo {pages} {096402} (\bibinfo {year} {2004})}\BibitemShut {NoStop}%
\bibitem [{\citenamefont {Kitaev}\ and\ \citenamefont
  {Preskill}(2006)}]{Kitaev2006}%
  \BibitemOpen
  \bibfield  {author} {\bibinfo {author} {\bibfnamefont {A.}~\bibnamefont
  {Kitaev}}\ and\ \bibinfo {author} {\bibfnamefont {J.}~\bibnamefont
  {Preskill}},\ }\bibfield  {title} {\bibinfo {title} {Topological entanglement
  entropy},\ }\href {https://doi.org/10.1103/PhysRevLett.96.110404} {\bibfield
  {journal} {\bibinfo  {journal} {Phys. Rev. Lett.}\ }\textbf {\bibinfo
  {volume} {96}},\ \bibinfo {pages} {110404} (\bibinfo {year}
  {2006})}\BibitemShut {NoStop}%
\bibitem [{\citenamefont {Levin}\ and\ \citenamefont {Wen}(2006)}]{Levin2006}%
  \BibitemOpen
  \bibfield  {author} {\bibinfo {author} {\bibfnamefont {M.}~\bibnamefont
  {Levin}}\ and\ \bibinfo {author} {\bibfnamefont {X.-G.}\ \bibnamefont
  {Wen}},\ }\bibfield  {title} {\bibinfo {title} {Detecting topological order
  in a ground state wave function},\ }\href
  {https://doi.org/10.1103/PhysRevLett.96.110405} {\bibfield  {journal}
  {\bibinfo  {journal} {Phys. Rev. Lett.}\ }\textbf {\bibinfo {volume} {96}},\
  \bibinfo {pages} {110405} (\bibinfo {year} {2006})}\BibitemShut {NoStop}%
\bibitem [{\citenamefont {Calabrese}\ and\ \citenamefont
  {Lefevre}(2008)}]{Calabrese2008entangle}%
  \BibitemOpen
  \bibfield  {author} {\bibinfo {author} {\bibfnamefont {P.}~\bibnamefont
  {Calabrese}}\ and\ \bibinfo {author} {\bibfnamefont {A.}~\bibnamefont
  {Lefevre}},\ }\bibfield  {title} {\bibinfo {title} {Entanglement spectrum in
  one-dimensional systems},\ }\href
  {https://doi.org/10.1103/PhysRevA.78.032329} {\bibfield  {journal} {\bibinfo
  {journal} {Phys. Rev. A}\ }\textbf {\bibinfo {volume} {78}},\ \bibinfo
  {pages} {032329} (\bibinfo {year} {2008})}\BibitemShut {NoStop}%
\bibitem [{\citenamefont {Fradkin}\ and\ \citenamefont
  {Moore}(2006)}]{Fradkin2006entangle}%
  \BibitemOpen
  \bibfield  {author} {\bibinfo {author} {\bibfnamefont {E.}~\bibnamefont
  {Fradkin}}\ and\ \bibinfo {author} {\bibfnamefont {J.~E.}\ \bibnamefont
  {Moore}},\ }\bibfield  {title} {\bibinfo {title} {Entanglement entropy of 2d
  conformal quantum critical points: Hearing the shape of a quantum drum},\
  }\href {https://doi.org/10.1103/PhysRevLett.97.050404} {\bibfield  {journal}
  {\bibinfo  {journal} {Phys. Rev. Lett.}\ }\textbf {\bibinfo {volume} {97}},\
  \bibinfo {pages} {050404} (\bibinfo {year} {2006})}\BibitemShut {NoStop}%
\bibitem [{\citenamefont {Nussinov}\ and\ \citenamefont
  {Ortiz}(2009{\natexlab{a}})}]{Nussinov2006}%
  \BibitemOpen
  \bibfield  {author} {\bibinfo {author} {\bibfnamefont {Z.}~\bibnamefont
  {Nussinov}}\ and\ \bibinfo {author} {\bibfnamefont {G.}~\bibnamefont
  {Ortiz}},\ }\bibfield  {title} {\bibinfo {title} {{Sufficient symmetry
  conditions for Topological Quantum Order}},\ }\href
  {https://doi.org/10.1073/pnas.0803726105} {\bibfield  {journal} {\bibinfo
  {journal} {Proc. Nat. Acad. Sci.}\ }\textbf {\bibinfo {volume} {106}},\
  \bibinfo {pages} {16944} (\bibinfo {year} {2009}{\natexlab{a}})}\BibitemShut
  {NoStop}%
\bibitem [{\citenamefont {Nussinov}\ and\ \citenamefont
  {Ortiz}(2009{\natexlab{b}})}]{Nussinov2009}%
  \BibitemOpen
  \bibfield  {author} {\bibinfo {author} {\bibfnamefont {Z.}~\bibnamefont
  {Nussinov}}\ and\ \bibinfo {author} {\bibfnamefont {G.}~\bibnamefont
  {Ortiz}},\ }\bibfield  {title} {\bibinfo {title} {{A symmetry principle for
  topological quantum order}},\ }\href
  {https://doi.org/10.1016/j.aop.2008.11.002} {\bibfield  {journal} {\bibinfo
  {journal} {Annals Phys.}\ }\textbf {\bibinfo {volume} {324}},\ \bibinfo
  {pages} {977} (\bibinfo {year} {2009}{\natexlab{b}})}\BibitemShut {NoStop}%
\bibitem [{\citenamefont {Casini}\ and\ \citenamefont
  {Huerta}(2007)}]{CASINI2007}%
  \BibitemOpen
  \bibfield  {author} {\bibinfo {author} {\bibfnamefont {H.}~\bibnamefont
  {Casini}}\ and\ \bibinfo {author} {\bibfnamefont {M.}~\bibnamefont
  {Huerta}},\ }\bibfield  {title} {\bibinfo {title} {Universal terms for the
  entanglement entropy in 2+1 dimensions},\ }\href
  {https://doi.org/https://doi.org/10.1016/j.nuclphysb.2006.12.012} {\bibfield
  {journal} {\bibinfo  {journal} {Nuclear Physics B}\ }\textbf {\bibinfo
  {volume} {764}},\ \bibinfo {pages} {183} (\bibinfo {year}
  {2007})}\BibitemShut {NoStop}%
\bibitem [{\citenamefont {Ji}\ and\ \citenamefont {Wen}(2019)}]{JiPRR2019}%
  \BibitemOpen
  \bibfield  {author} {\bibinfo {author} {\bibfnamefont {W.}~\bibnamefont
  {Ji}}\ and\ \bibinfo {author} {\bibfnamefont {X.-G.}\ \bibnamefont {Wen}},\
  }\bibfield  {title} {\bibinfo {title} {Noninvertible anomalies and
  mapping-class-group transformation of anomalous partition functions},\ }\href
  {https://doi.org/10.1103/PhysRevResearch.1.033054} {\bibfield  {journal}
  {\bibinfo  {journal} {Phys. Rev. Research}\ }\textbf {\bibinfo {volume}
  {1}},\ \bibinfo {pages} {033054} (\bibinfo {year} {2019})}\BibitemShut
  {NoStop}%
\bibitem [{\citenamefont {Ji}\ and\ \citenamefont
  {Wen}(2020)}]{ji2019categorical}%
  \BibitemOpen
  \bibfield  {author} {\bibinfo {author} {\bibfnamefont {W.}~\bibnamefont
  {Ji}}\ and\ \bibinfo {author} {\bibfnamefont {X.-G.}\ \bibnamefont {Wen}},\
  }\bibfield  {title} {\bibinfo {title} {Categorical symmetry and noninvertible
  anomaly in symmetry-breaking and topological phase transitions},\ }\href
  {https://doi.org/10.1103/PhysRevResearch.2.033417} {\bibfield  {journal}
  {\bibinfo  {journal} {Phys. Rev. Research}\ }\textbf {\bibinfo {volume}
  {2}},\ \bibinfo {pages} {033417} (\bibinfo {year} {2020})}\BibitemShut
  {NoStop}%
\bibitem [{\citenamefont {Kong}\ \emph {et~al.}(2020)\citenamefont {Kong},
  \citenamefont {Lan}, \citenamefont {Wen}, \citenamefont {Zhang},\ and\
  \citenamefont {Zheng}}]{kong2020algebraic}%
  \BibitemOpen
  \bibfield  {author} {\bibinfo {author} {\bibfnamefont {L.}~\bibnamefont
  {Kong}}, \bibinfo {author} {\bibfnamefont {T.}~\bibnamefont {Lan}}, \bibinfo
  {author} {\bibfnamefont {X.-G.}\ \bibnamefont {Wen}}, \bibinfo {author}
  {\bibfnamefont {Z.-H.}\ \bibnamefont {Zhang}},\ and\ \bibinfo {author}
  {\bibfnamefont {H.}~\bibnamefont {Zheng}},\ }\bibfield  {title} {\bibinfo
  {title} {Algebraic higher symmetry and categorical symmetry: A holographic
  and entanglement view of symmetry},\ }\href
  {https://doi.org/10.1103/PhysRevResearch.2.043086} {\bibfield  {journal}
  {\bibinfo  {journal} {Phys. Rev. Research}\ }\textbf {\bibinfo {volume}
  {2}},\ \bibinfo {pages} {043086} (\bibinfo {year} {2020})}\BibitemShut
  {NoStop}%
\bibitem [{\citenamefont {Wu}\ \emph {et~al.}(2021{\natexlab{a}})\citenamefont
  {Wu}, \citenamefont {Ji},\ and\ \citenamefont {Xu}}]{XCWu2020}%
  \BibitemOpen
  \bibfield  {author} {\bibinfo {author} {\bibfnamefont {X.-C.}\ \bibnamefont
  {Wu}}, \bibinfo {author} {\bibfnamefont {W.}~\bibnamefont {Ji}},\ and\
  \bibinfo {author} {\bibfnamefont {C.}~\bibnamefont {Xu}},\ }\bibfield
  {title} {\bibinfo {title} {Categorical symmetries at criticality},\ }\href
  {https://doi.org/10.1088/1742-5468/ac08fe} {\bibfield  {journal} {\bibinfo
  {journal} {Journal of Statistical Mechanics: Theory and Experiment}\ }\textbf
  {\bibinfo {volume} {2021}},\ \bibinfo {pages} {073101} (\bibinfo {year}
  {2021}{\natexlab{a}})}\BibitemShut {NoStop}%
\bibitem [{\citenamefont {Wu}\ \emph {et~al.}(2021{\natexlab{b}})\citenamefont
  {Wu}, \citenamefont {Jian},\ and\ \citenamefont {Xu}}]{XCWu2021}%
  \BibitemOpen
  \bibfield  {author} {\bibinfo {author} {\bibfnamefont {X.-C.}\ \bibnamefont
  {Wu}}, \bibinfo {author} {\bibfnamefont {C.-M.}\ \bibnamefont {Jian}},\ and\
  \bibinfo {author} {\bibfnamefont {C.}~\bibnamefont {Xu}},\ }\bibfield
  {title} {\bibinfo {title} {{Universal Features of Higher-Form Symmetries at
  Phase Transitions}},\ }\href {https://doi.org/10.21468/SciPostPhys.11.2.033}
  {\bibfield  {journal} {\bibinfo  {journal} {SciPost Phys.}\ }\textbf
  {\bibinfo {volume} {11}},\ \bibinfo {pages} {33} (\bibinfo {year}
  {2021}{\natexlab{b}})}\BibitemShut {NoStop}%
\bibitem [{\citenamefont {Zhao}\ \emph
  {et~al.}(2022{\natexlab{a}})\citenamefont {Zhao}, \citenamefont {Wang},
  \citenamefont {Yan}, \citenamefont {Cheng},\ and\ \citenamefont
  {Meng}}]{JRZhao2021}%
  \BibitemOpen
  \bibfield  {author} {\bibinfo {author} {\bibfnamefont {J.}~\bibnamefont
  {Zhao}}, \bibinfo {author} {\bibfnamefont {Y.-C.}\ \bibnamefont {Wang}},
  \bibinfo {author} {\bibfnamefont {Z.}~\bibnamefont {Yan}}, \bibinfo {author}
  {\bibfnamefont {M.}~\bibnamefont {Cheng}},\ and\ \bibinfo {author}
  {\bibfnamefont {Z.~Y.}\ \bibnamefont {Meng}},\ }\bibfield  {title} {\bibinfo
  {title} {Scaling of entanglement entropy at deconfined quantum criticality},\
  }\href {https://doi.org/10.1103/PhysRevLett.128.010601} {\bibfield  {journal}
  {\bibinfo  {journal} {Phys. Rev. Lett.}\ }\textbf {\bibinfo {volume} {128}},\
  \bibinfo {pages} {010601} (\bibinfo {year} {2022}{\natexlab{a}})}\BibitemShut
  {NoStop}%
\bibitem [{\citenamefont {Zhao}\ \emph
  {et~al.}(2022{\natexlab{b}})\citenamefont {Zhao}, \citenamefont {Chen},
  \citenamefont {Wang}, \citenamefont {Yan}, \citenamefont {Cheng},\ and\
  \citenamefont {Meng}}]{JRZhao2022}%
  \BibitemOpen
  \bibfield  {author} {\bibinfo {author} {\bibfnamefont {J.}~\bibnamefont
  {Zhao}}, \bibinfo {author} {\bibfnamefont {B.-B.}\ \bibnamefont {Chen}},
  \bibinfo {author} {\bibfnamefont {Y.-C.}\ \bibnamefont {Wang}}, \bibinfo
  {author} {\bibfnamefont {Z.}~\bibnamefont {Yan}}, \bibinfo {author}
  {\bibfnamefont {M.}~\bibnamefont {Cheng}},\ and\ \bibinfo {author}
  {\bibfnamefont {Z.~Y.}\ \bibnamefont {Meng}},\ }\bibfield  {title} {\bibinfo
  {title} {Measuring r{\'e}nyi entanglement entropy with high efficiency and
  precision in quantum monte carlo simulations},\ }\href
  {https://doi.org/10.1038/s41535-022-00476-0} {\bibfield  {journal} {\bibinfo
  {journal} {npj Quantum Materials}\ }\textbf {\bibinfo {volume} {7}},\
  \bibinfo {pages} {69} (\bibinfo {year} {2022}{\natexlab{b}})}\BibitemShut
  {NoStop}%
\bibitem [{\citenamefont {Wang}\ \emph {et~al.}(2022)\citenamefont {Wang},
  \citenamefont {Ma}, \citenamefont {Cheng},\ and\ \citenamefont
  {Meng}}]{10.21468/SciPostPhys.13.6.123}%
  \BibitemOpen
  \bibfield  {author} {\bibinfo {author} {\bibfnamefont {Y.-C.}\ \bibnamefont
  {Wang}}, \bibinfo {author} {\bibfnamefont {N.}~\bibnamefont {Ma}}, \bibinfo
  {author} {\bibfnamefont {M.}~\bibnamefont {Cheng}},\ and\ \bibinfo {author}
  {\bibfnamefont {Z.~Y.}\ \bibnamefont {Meng}},\ }\bibfield  {title} {\bibinfo
  {title} {{Scaling of the disorder operator at deconfined quantum
  criticality}},\ }\href {https://doi.org/10.21468/SciPostPhys.13.6.123}
  {\bibfield  {journal} {\bibinfo  {journal} {SciPost Phys.}\ }\textbf
  {\bibinfo {volume} {13}},\ \bibinfo {pages} {123} (\bibinfo {year}
  {2022})}\BibitemShut {NoStop}%
\bibitem [{\citenamefont {Pollmann}\ \emph {et~al.}(2010)\citenamefont
  {Pollmann}, \citenamefont {Turner}, \citenamefont {Berg},\ and\ \citenamefont
  {Oshikawa}}]{Pollmann2010entangle}%
  \BibitemOpen
  \bibfield  {author} {\bibinfo {author} {\bibfnamefont {F.}~\bibnamefont
  {Pollmann}}, \bibinfo {author} {\bibfnamefont {A.~M.}\ \bibnamefont
  {Turner}}, \bibinfo {author} {\bibfnamefont {E.}~\bibnamefont {Berg}},\ and\
  \bibinfo {author} {\bibfnamefont {M.}~\bibnamefont {Oshikawa}},\ }\bibfield
  {title} {\bibinfo {title} {Entanglement spectrum of a topological phase in
  one dimension},\ }\href {https://doi.org/10.1103/PhysRevB.81.064439}
  {\bibfield  {journal} {\bibinfo  {journal} {Phys. Rev. B}\ }\textbf {\bibinfo
  {volume} {81}},\ \bibinfo {pages} {064439} (\bibinfo {year}
  {2010})}\BibitemShut {NoStop}%
\bibitem [{\citenamefont {Fidkowski}(2010)}]{Fidkowski2010}%
  \BibitemOpen
  \bibfield  {author} {\bibinfo {author} {\bibfnamefont {L.}~\bibnamefont
  {Fidkowski}},\ }\bibfield  {title} {\bibinfo {title} {Entanglement spectrum
  of topological insulators and superconductors},\ }\href
  {https://doi.org/10.1103/PhysRevLett.104.130502} {\bibfield  {journal}
  {\bibinfo  {journal} {Phys. Rev. Lett.}\ }\textbf {\bibinfo {volume} {104}},\
  \bibinfo {pages} {130502} (\bibinfo {year} {2010})}\BibitemShut {NoStop}%
\bibitem [{\citenamefont {Yao}\ and\ \citenamefont {Qi}(2010)}]{Yao2010}%
  \BibitemOpen
  \bibfield  {author} {\bibinfo {author} {\bibfnamefont {H.}~\bibnamefont
  {Yao}}\ and\ \bibinfo {author} {\bibfnamefont {X.-L.}\ \bibnamefont {Qi}},\
  }\bibfield  {title} {\bibinfo {title} {Entanglement entropy and entanglement
  spectrum of the kitaev model},\ }\href
  {https://doi.org/10.1103/PhysRevLett.105.080501} {\bibfield  {journal}
  {\bibinfo  {journal} {Phys. Rev. Lett.}\ }\textbf {\bibinfo {volume} {105}},\
  \bibinfo {pages} {080501} (\bibinfo {year} {2010})}\BibitemShut {NoStop}%
\bibitem [{\citenamefont {Qi}\ \emph {et~al.}(2012)\citenamefont {Qi},
  \citenamefont {Katsura},\ and\ \citenamefont {Ludwig}}]{XLQi2012}%
  \BibitemOpen
  \bibfield  {author} {\bibinfo {author} {\bibfnamefont {X.-L.}\ \bibnamefont
  {Qi}}, \bibinfo {author} {\bibfnamefont {H.}~\bibnamefont {Katsura}},\ and\
  \bibinfo {author} {\bibfnamefont {A.~W.~W.}\ \bibnamefont {Ludwig}},\
  }\bibfield  {title} {\bibinfo {title} {General relationship between the
  entanglement spectrum and the edge state spectrum of topological quantum
  states},\ }\href {https://doi.org/10.1103/PhysRevLett.108.196402} {\bibfield
  {journal} {\bibinfo  {journal} {Phys. Rev. Lett.}\ }\textbf {\bibinfo
  {volume} {108}},\ \bibinfo {pages} {196402} (\bibinfo {year}
  {2012})}\BibitemShut {NoStop}%
\bibitem [{\citenamefont {Canovi}\ \emph {et~al.}(2014)\citenamefont {Canovi},
  \citenamefont {Ercolessi}, \citenamefont {Naldesi}, \citenamefont {Taddia},\
  and\ \citenamefont {Vodola}}]{Canovi2014}%
  \BibitemOpen
  \bibfield  {author} {\bibinfo {author} {\bibfnamefont {E.}~\bibnamefont
  {Canovi}}, \bibinfo {author} {\bibfnamefont {E.}~\bibnamefont {Ercolessi}},
  \bibinfo {author} {\bibfnamefont {P.}~\bibnamefont {Naldesi}}, \bibinfo
  {author} {\bibfnamefont {L.}~\bibnamefont {Taddia}},\ and\ \bibinfo {author}
  {\bibfnamefont {D.}~\bibnamefont {Vodola}},\ }\bibfield  {title} {\bibinfo
  {title} {Dynamics of entanglement entropy and entanglement spectrum crossing
  a quantum phase transition},\ }\href
  {https://doi.org/10.1103/PhysRevB.89.104303} {\bibfield  {journal} {\bibinfo
  {journal} {Phys. Rev. B}\ }\textbf {\bibinfo {volume} {89}},\ \bibinfo
  {pages} {104303} (\bibinfo {year} {2014})}\BibitemShut {NoStop}%
\bibitem [{\citenamefont {Luitz}\ \emph
  {et~al.}(2014{\natexlab{a}})\citenamefont {Luitz}, \citenamefont {Alet},\
  and\ \citenamefont {Laflorencie}}]{LuitzPRL2014}%
  \BibitemOpen
  \bibfield  {author} {\bibinfo {author} {\bibfnamefont {D.~J.}\ \bibnamefont
  {Luitz}}, \bibinfo {author} {\bibfnamefont {F.}~\bibnamefont {Alet}},\ and\
  \bibinfo {author} {\bibfnamefont {N.}~\bibnamefont {Laflorencie}},\
  }\bibfield  {title} {\bibinfo {title} {Universal behavior beyond
  multifractality in quantum many-body systems},\ }\href
  {https://doi.org/10.1103/PhysRevLett.112.057203} {\bibfield  {journal}
  {\bibinfo  {journal} {Phys. Rev. Lett.}\ }\textbf {\bibinfo {volume} {112}},\
  \bibinfo {pages} {057203} (\bibinfo {year} {2014}{\natexlab{a}})}\BibitemShut
  {NoStop}%
\bibitem [{\citenamefont {Luitz}\ \emph
  {et~al.}(2014{\natexlab{b}})\citenamefont {Luitz}, \citenamefont {Alet},\
  and\ \citenamefont {Laflorencie}}]{LuitzPRB2014}%
  \BibitemOpen
  \bibfield  {author} {\bibinfo {author} {\bibfnamefont {D.~J.}\ \bibnamefont
  {Luitz}}, \bibinfo {author} {\bibfnamefont {F.}~\bibnamefont {Alet}},\ and\
  \bibinfo {author} {\bibfnamefont {N.}~\bibnamefont {Laflorencie}},\
  }\bibfield  {title} {\bibinfo {title} {Shannon-r\'enyi entropies and
  participation spectra across three-dimensional $o(3)$ criticality},\ }\href
  {https://doi.org/10.1103/PhysRevB.89.165106} {\bibfield  {journal} {\bibinfo
  {journal} {Phys. Rev. B}\ }\textbf {\bibinfo {volume} {89}},\ \bibinfo
  {pages} {165106} (\bibinfo {year} {2014}{\natexlab{b}})}\BibitemShut
  {NoStop}%
\bibitem [{\citenamefont {Luitz}\ \emph
  {et~al.}(2014{\natexlab{c}})\citenamefont {Luitz}, \citenamefont
  {Laflorencie},\ and\ \citenamefont {Alet}}]{LuitzIOP2014}%
  \BibitemOpen
  \bibfield  {author} {\bibinfo {author} {\bibfnamefont {D.~J.}\ \bibnamefont
  {Luitz}}, \bibinfo {author} {\bibfnamefont {N.}~\bibnamefont {Laflorencie}},\
  and\ \bibinfo {author} {\bibfnamefont {F.}~\bibnamefont {Alet}},\ }\bibfield
  {title} {\bibinfo {title} {Participation spectroscopy and entanglement
  hamiltonian of quantum spin models},\ }\href
  {https://doi.org/10.1088/1742-5468/2014/08/p08007} {\bibfield  {journal}
  {\bibinfo  {journal} {Journal of Statistical Mechanics: Theory and
  Experiment}\ }\textbf {\bibinfo {volume} {2014}},\ \bibinfo {pages} {P08007}
  (\bibinfo {year} {2014}{\natexlab{c}})}\BibitemShut {NoStop}%
\bibitem [{\citenamefont {Chung}\ \emph {et~al.}(2014)\citenamefont {Chung},
  \citenamefont {Bonnes}, \citenamefont {Chen},\ and\ \citenamefont
  {L\"auchli}}]{Chung2014}%
  \BibitemOpen
  \bibfield  {author} {\bibinfo {author} {\bibfnamefont {C.-M.}\ \bibnamefont
  {Chung}}, \bibinfo {author} {\bibfnamefont {L.}~\bibnamefont {Bonnes}},
  \bibinfo {author} {\bibfnamefont {P.}~\bibnamefont {Chen}},\ and\ \bibinfo
  {author} {\bibfnamefont {A.~M.}\ \bibnamefont {L\"auchli}},\ }\bibfield
  {title} {\bibinfo {title} {Entanglement spectroscopy using quantum monte
  carlo},\ }\href {https://doi.org/10.1103/PhysRevB.89.195147} {\bibfield
  {journal} {\bibinfo  {journal} {Phys. Rev. B}\ }\textbf {\bibinfo {volume}
  {89}},\ \bibinfo {pages} {195147} (\bibinfo {year} {2014})}\BibitemShut
  {NoStop}%
\bibitem [{\citenamefont {Pichler}\ \emph {et~al.}(2016)\citenamefont
  {Pichler}, \citenamefont {Zhu}, \citenamefont {Seif}, \citenamefont
  {Zoller},\ and\ \citenamefont {Hafezi}}]{Pichler2016}%
  \BibitemOpen
  \bibfield  {author} {\bibinfo {author} {\bibfnamefont {H.}~\bibnamefont
  {Pichler}}, \bibinfo {author} {\bibfnamefont {G.}~\bibnamefont {Zhu}},
  \bibinfo {author} {\bibfnamefont {A.}~\bibnamefont {Seif}}, \bibinfo {author}
  {\bibfnamefont {P.}~\bibnamefont {Zoller}},\ and\ \bibinfo {author}
  {\bibfnamefont {M.}~\bibnamefont {Hafezi}},\ }\bibfield  {title} {\bibinfo
  {title} {Measurement protocol for the entanglement spectrum of cold atoms},\
  }\href {https://doi.org/10.1103/PhysRevX.6.041033} {\bibfield  {journal}
  {\bibinfo  {journal} {Phys. Rev. X}\ }\textbf {\bibinfo {volume} {6}},\
  \bibinfo {pages} {041033} (\bibinfo {year} {2016})}\BibitemShut {NoStop}%
\bibitem [{\citenamefont {Cirac}\ \emph
  {et~al.}(2011{\natexlab{a}})\citenamefont {Cirac}, \citenamefont {Poilblanc},
  \citenamefont {Schuch},\ and\ \citenamefont {Verstraete}}]{Cirac2011}%
  \BibitemOpen
  \bibfield  {author} {\bibinfo {author} {\bibfnamefont {J.~I.}\ \bibnamefont
  {Cirac}}, \bibinfo {author} {\bibfnamefont {D.}~\bibnamefont {Poilblanc}},
  \bibinfo {author} {\bibfnamefont {N.}~\bibnamefont {Schuch}},\ and\ \bibinfo
  {author} {\bibfnamefont {F.}~\bibnamefont {Verstraete}},\ }\bibfield  {title}
  {\bibinfo {title} {Entanglement spectrum and boundary theories with projected
  entangled-pair states},\ }\href {https://doi.org/10.1103/PhysRevB.83.245134}
  {\bibfield  {journal} {\bibinfo  {journal} {Phys. Rev. B}\ }\textbf {\bibinfo
  {volume} {83}},\ \bibinfo {pages} {245134} (\bibinfo {year}
  {2011}{\natexlab{a}})}\BibitemShut {NoStop}%
\bibitem [{\citenamefont {Stojanovi\ifmmode~\acute{c}\else
  \'{c}\fi{}}(2020)}]{Stoj2020}%
  \BibitemOpen
  \bibfield  {author} {\bibinfo {author} {\bibfnamefont {V.~M.}\ \bibnamefont
  {Stojanovi\ifmmode~\acute{c}\else \'{c}\fi{}}},\ }\bibfield  {title}
  {\bibinfo {title} {Entanglement-spectrum characterization of ground-state
  nonanalyticities in coupled excitation-phonon models},\ }\href
  {https://doi.org/10.1103/PhysRevB.101.134301} {\bibfield  {journal} {\bibinfo
   {journal} {Phys. Rev. B}\ }\textbf {\bibinfo {volume} {101}},\ \bibinfo
  {pages} {134301} (\bibinfo {year} {2020})}\BibitemShut {NoStop}%
\bibitem [{\citenamefont {Guo}(2021)}]{guo2021entanglement}%
  \BibitemOpen
  \bibfield  {author} {\bibinfo {author} {\bibfnamefont {W.-z.}\ \bibnamefont
  {Guo}},\ }\bibfield  {title} {\bibinfo {title} {Entanglement spectrum of
  geometric states},\ }\href
  {https://link.springer.com/article/10.1007/JHEP02(2021)085} {\bibfield
  {journal} {\bibinfo  {journal} {Journal of High Energy Physics}\ }\textbf
  {\bibinfo {volume} {2021}},\ \bibinfo {pages} {1} (\bibinfo {year}
  {2021})}\BibitemShut {NoStop}%
\bibitem [{\citenamefont {Grover}(2013)}]{Grover2013}%
  \BibitemOpen
  \bibfield  {author} {\bibinfo {author} {\bibfnamefont {T.}~\bibnamefont
  {Grover}},\ }\bibfield  {title} {\bibinfo {title} {Entanglement of
  interacting fermions in quantum monte carlo calculations},\ }\href
  {https://doi.org/10.1103/PhysRevLett.111.130402} {\bibfield  {journal}
  {\bibinfo  {journal} {Phys. Rev. Lett.}\ }\textbf {\bibinfo {volume} {111}},\
  \bibinfo {pages} {130402} (\bibinfo {year} {2013})}\BibitemShut {NoStop}%
\bibitem [{\citenamefont {Assaad}\ \emph {et~al.}(2014)\citenamefont {Assaad},
  \citenamefont {Lang},\ and\ \citenamefont {Parisen~Toldin}}]{Assaad2014}%
  \BibitemOpen
  \bibfield  {author} {\bibinfo {author} {\bibfnamefont {F.~F.}\ \bibnamefont
  {Assaad}}, \bibinfo {author} {\bibfnamefont {T.~C.}\ \bibnamefont {Lang}},\
  and\ \bibinfo {author} {\bibfnamefont {F.}~\bibnamefont {Parisen~Toldin}},\
  }\bibfield  {title} {\bibinfo {title} {Entanglement spectra of interacting
  fermions in quantum monte carlo simulations},\ }\href
  {https://doi.org/10.1103/PhysRevB.89.125121} {\bibfield  {journal} {\bibinfo
  {journal} {Phys. Rev. B}\ }\textbf {\bibinfo {volume} {89}},\ \bibinfo
  {pages} {125121} (\bibinfo {year} {2014})}\BibitemShut {NoStop}%
\bibitem [{\citenamefont {Assaad}(2015)}]{Assaad2015}%
  \BibitemOpen
  \bibfield  {author} {\bibinfo {author} {\bibfnamefont {F.~F.}\ \bibnamefont
  {Assaad}},\ }\bibfield  {title} {\bibinfo {title} {Stable quantum monte carlo
  simulations for entanglement spectra of interacting fermions},\ }\href
  {https://doi.org/10.1103/PhysRevB.91.125146} {\bibfield  {journal} {\bibinfo
  {journal} {Phys. Rev. B}\ }\textbf {\bibinfo {volume} {91}},\ \bibinfo
  {pages} {125146} (\bibinfo {year} {2015})}\BibitemShut {NoStop}%
\bibitem [{\citenamefont {Parisen~Toldin}\ and\ \citenamefont
  {Assaad}(2018)}]{Parisen2018}%
  \BibitemOpen
  \bibfield  {author} {\bibinfo {author} {\bibfnamefont {F.}~\bibnamefont
  {Parisen~Toldin}}\ and\ \bibinfo {author} {\bibfnamefont {F.~F.}\
  \bibnamefont {Assaad}},\ }\bibfield  {title} {\bibinfo {title} {Entanglement
  hamiltonian of interacting fermionic models},\ }\href
  {https://doi.org/10.1103/PhysRevLett.121.200602} {\bibfield  {journal}
  {\bibinfo  {journal} {Phys. Rev. Lett.}\ }\textbf {\bibinfo {volume} {121}},\
  \bibinfo {pages} {200602} (\bibinfo {year} {2018})}\BibitemShut {NoStop}%
\bibitem [{\citenamefont {Yu}\ \emph {et~al.}(2022)\citenamefont {Yu},
  \citenamefont {Huang}, \citenamefont {Song}, \citenamefont {Xu},
  \citenamefont {Ding},\ and\ \citenamefont {Zhang}}]{PhysRevLett.129.210601}%
  \BibitemOpen
  \bibfield  {author} {\bibinfo {author} {\bibfnamefont {X.-J.}\ \bibnamefont
  {Yu}}, \bibinfo {author} {\bibfnamefont {R.-Z.}\ \bibnamefont {Huang}},
  \bibinfo {author} {\bibfnamefont {H.-H.}\ \bibnamefont {Song}}, \bibinfo
  {author} {\bibfnamefont {L.}~\bibnamefont {Xu}}, \bibinfo {author}
  {\bibfnamefont {C.}~\bibnamefont {Ding}},\ and\ \bibinfo {author}
  {\bibfnamefont {L.}~\bibnamefont {Zhang}},\ }\bibfield  {title} {\bibinfo
  {title} {Conformal boundary conditions of symmetry-enriched quantum critical
  spin chains},\ }\href {https://doi.org/10.1103/PhysRevLett.129.210601}
  {\bibfield  {journal} {\bibinfo  {journal} {Phys. Rev. Lett.}\ }\textbf
  {\bibinfo {volume} {129}},\ \bibinfo {pages} {210601} (\bibinfo {year}
  {2022})}\BibitemShut {NoStop}%
\bibitem [{\citenamefont {Moradi}\ and\ \citenamefont
  {Abouie}(2016)}]{Moradi_2016}%
  \BibitemOpen
  \bibfield  {author} {\bibinfo {author} {\bibfnamefont {Z.}~\bibnamefont
  {Moradi}}\ and\ \bibinfo {author} {\bibfnamefont {J.}~\bibnamefont
  {Abouie}},\ }\bibfield  {title} {\bibinfo {title} {Entanglement spectrum of
  fermionic bilayer honeycomb lattice: Hofstadter butterfly},\ }\href
  {https://doi.org/10.1088/1742-5468/2016/11/113101} {\bibfield  {journal}
  {\bibinfo  {journal} {Journal of Statistical Mechanics: Theory and
  Experiment}\ }\textbf {\bibinfo {volume} {2016}},\ \bibinfo {pages} {113101}
  (\bibinfo {year} {2016})}\BibitemShut {NoStop}%
\bibitem [{\citenamefont {Haldane}(1983{\natexlab{a}})}]{haldaneContinuum1983}%
  \BibitemOpen
  \bibfield  {author} {\bibinfo {author} {\bibfnamefont {F.}~\bibnamefont
  {Haldane}},\ }\bibfield  {title} {\bibinfo {title} {Continuum dynamics of the
  1-d heisenberg antiferromagnet: Identification with the o(3) nonlinear sigma
  model},\ }\href
  {https://doi.org/https://doi.org/10.1016/0375-9601(83)90631-X} {\bibfield
  {journal} {\bibinfo  {journal} {Physics Letters A}\ }\textbf {\bibinfo
  {volume} {93}},\ \bibinfo {pages} {464} (\bibinfo {year}
  {1983}{\natexlab{a}})}\BibitemShut {NoStop}%
\bibitem [{\citenamefont {Haldane}(1983{\natexlab{b}})}]{haldaneNonlinear1983}%
  \BibitemOpen
  \bibfield  {author} {\bibinfo {author} {\bibfnamefont {F.~D.~M.}\
  \bibnamefont {Haldane}},\ }\bibfield  {title} {\bibinfo {title} {Nonlinear
  field theory of large-spin heisenberg antiferromagnets: Semiclassically
  quantized solitons of the one-dimensional easy-axis n\'eel state},\ }\href
  {https://doi.org/10.1103/PhysRevLett.50.1153} {\bibfield  {journal} {\bibinfo
   {journal} {Phys. Rev. Lett.}\ }\textbf {\bibinfo {volume} {50}},\ \bibinfo
  {pages} {1153} (\bibinfo {year} {1983}{\natexlab{b}})}\BibitemShut {NoStop}%
\bibitem [{\citenamefont {Li}\ and\ \citenamefont
  {Haldane}(2008)}]{Li2008entangle}%
  \BibitemOpen
  \bibfield  {author} {\bibinfo {author} {\bibfnamefont {H.}~\bibnamefont
  {Li}}\ and\ \bibinfo {author} {\bibfnamefont {F.~D.~M.}\ \bibnamefont
  {Haldane}},\ }\bibfield  {title} {\bibinfo {title} {Entanglement spectrum as
  a generalization of entanglement entropy: Identification of topological order
  in non-abelian fractional quantum hall effect states},\ }\href
  {https://doi.org/10.1103/PhysRevLett.101.010504} {\bibfield  {journal}
  {\bibinfo  {journal} {Phys. Rev. Lett.}\ }\textbf {\bibinfo {volume} {101}},\
  \bibinfo {pages} {010504} (\bibinfo {year} {2008})}\BibitemShut {NoStop}%
\bibitem [{\citenamefont {Poilblanc}(2010)}]{Poilblanc2010entanglement}%
  \BibitemOpen
  \bibfield  {author} {\bibinfo {author} {\bibfnamefont {D.}~\bibnamefont
  {Poilblanc}},\ }\bibfield  {title} {\bibinfo {title} {Entanglement spectra of
  quantum heisenberg ladders},\ }\href
  {https://doi.org/10.1103/PhysRevLett.105.077202} {\bibfield  {journal}
  {\bibinfo  {journal} {Phys. Rev. Lett.}\ }\textbf {\bibinfo {volume} {105}},\
  \bibinfo {pages} {077202} (\bibinfo {year} {2010})}\BibitemShut {NoStop}%
\bibitem [{\citenamefont {Wang}\ \emph {et~al.}(2006)\citenamefont {Wang},
  \citenamefont {Beach},\ and\ \citenamefont {Sandvik}}]{Wang2006bilayer}%
  \BibitemOpen
  \bibfield  {author} {\bibinfo {author} {\bibfnamefont {L.}~\bibnamefont
  {Wang}}, \bibinfo {author} {\bibfnamefont {K.~S.~D.}\ \bibnamefont {Beach}},\
  and\ \bibinfo {author} {\bibfnamefont {A.~W.}\ \bibnamefont {Sandvik}},\
  }\bibfield  {title} {\bibinfo {title} {High-precision finite-size scaling
  analysis of the quantum-critical point of heisenberg antiferromagnetic
  bilayers},\ }\href {https://doi.org/10.1103/PhysRevB.73.014431} {\bibfield
  {journal} {\bibinfo  {journal} {Phys. Rev. B}\ }\textbf {\bibinfo {volume}
  {73}},\ \bibinfo {pages} {014431} (\bibinfo {year} {2006})}\BibitemShut
  {NoStop}%
\bibitem [{\citenamefont {Loh\"ofer}\ \emph {et~al.}(2015)\citenamefont
  {Loh\"ofer}, \citenamefont {Coletta}, \citenamefont {Joshi}, \citenamefont
  {Assaad}, \citenamefont {Vojta}, \citenamefont {Wessel},\ and\ \citenamefont
  {Mila}}]{Lohoefer2015}%
  \BibitemOpen
  \bibfield  {author} {\bibinfo {author} {\bibfnamefont {M.}~\bibnamefont
  {Loh\"ofer}}, \bibinfo {author} {\bibfnamefont {T.}~\bibnamefont {Coletta}},
  \bibinfo {author} {\bibfnamefont {D.~G.}\ \bibnamefont {Joshi}}, \bibinfo
  {author} {\bibfnamefont {F.~F.}\ \bibnamefont {Assaad}}, \bibinfo {author}
  {\bibfnamefont {M.}~\bibnamefont {Vojta}}, \bibinfo {author} {\bibfnamefont
  {S.}~\bibnamefont {Wessel}},\ and\ \bibinfo {author} {\bibfnamefont
  {F.}~\bibnamefont {Mila}},\ }\bibfield  {title} {\bibinfo {title} {Dynamical
  structure factors and excitation modes of the bilayer heisenberg model},\
  }\href {https://doi.org/10.1103/PhysRevB.92.245137} {\bibfield  {journal}
  {\bibinfo  {journal} {Phys. Rev. B}\ }\textbf {\bibinfo {volume} {92}},\
  \bibinfo {pages} {245137} (\bibinfo {year} {2015})}\BibitemShut {NoStop}%
\bibitem [{\citenamefont {Alba}\ \emph {et~al.}(2012)\citenamefont {Alba},
  \citenamefont {Haque},\ and\ \citenamefont
  {Läuchli}}]{alba2012entanglement}%
  \BibitemOpen
  \bibfield  {author} {\bibinfo {author} {\bibfnamefont {V.}~\bibnamefont
  {Alba}}, \bibinfo {author} {\bibfnamefont {M.}~\bibnamefont {Haque}},\ and\
  \bibinfo {author} {\bibfnamefont {A.~M.}\ \bibnamefont {Läuchli}},\
  }\bibfield  {title} {\bibinfo {title} {Entanglement spectrum of the
  heisenberg xxz chain near the ferromagnetic point},\ }\href
  {https://doi.org/10.1088/1742-5468/2012/08/P08011} {\bibfield  {journal}
  {\bibinfo  {journal} {Journal of Statistical Mechanics: Theory and
  Experiment}\ }\textbf {\bibinfo {volume} {2012}},\ \bibinfo {pages} {P08011}
  (\bibinfo {year} {2012})}\BibitemShut {NoStop}%
\bibitem [{\citenamefont {Alba}\ \emph {et~al.}(2013)\citenamefont {Alba},
  \citenamefont {Haque},\ and\ \citenamefont
  {L\"auchli}}]{PhysRevLett.110.260403}%
  \BibitemOpen
  \bibfield  {author} {\bibinfo {author} {\bibfnamefont {V.}~\bibnamefont
  {Alba}}, \bibinfo {author} {\bibfnamefont {M.}~\bibnamefont {Haque}},\ and\
  \bibinfo {author} {\bibfnamefont {A.~M.}\ \bibnamefont {L\"auchli}},\
  }\bibfield  {title} {\bibinfo {title} {Entanglement spectrum of the
  two-dimensional bose-hubbard model},\ }\href
  {https://doi.org/10.1103/PhysRevLett.110.260403} {\bibfield  {journal}
  {\bibinfo  {journal} {Phys. Rev. Lett.}\ }\textbf {\bibinfo {volume} {110}},\
  \bibinfo {pages} {260403} (\bibinfo {year} {2013})}\BibitemShut {NoStop}%
\bibitem [{\citenamefont {Zhu}\ \emph {et~al.}(2020)\citenamefont {Zhu},
  \citenamefont {Huang}, \citenamefont {He},\ and\ \citenamefont
  {Wen}}]{zhu2020entanglement}%
  \BibitemOpen
  \bibfield  {author} {\bibinfo {author} {\bibfnamefont {W.}~\bibnamefont
  {Zhu}}, \bibinfo {author} {\bibfnamefont {Z.}~\bibnamefont {Huang}}, \bibinfo
  {author} {\bibfnamefont {Y.-C.}\ \bibnamefont {He}},\ and\ \bibinfo {author}
  {\bibfnamefont {X.}~\bibnamefont {Wen}},\ }\bibfield  {title} {\bibinfo
  {title} {Entanglement hamiltonian of many-body dynamics in strongly
  correlated systems},\ }\href {https://doi.org/10.1103/PhysRevLett.124.100605}
  {\bibfield  {journal} {\bibinfo  {journal} {Phys. Rev. Lett.}\ }\textbf
  {\bibinfo {volume} {124}},\ \bibinfo {pages} {100605} (\bibinfo {year}
  {2020})}\BibitemShut {NoStop}%
\bibitem [{\citenamefont {Zhu}\ \emph {et~al.}(2019)\citenamefont {Zhu},
  \citenamefont {Huang},\ and\ \citenamefont {He}}]{zhu2019reconstructing}%
  \BibitemOpen
  \bibfield  {author} {\bibinfo {author} {\bibfnamefont {W.}~\bibnamefont
  {Zhu}}, \bibinfo {author} {\bibfnamefont {Z.}~\bibnamefont {Huang}},\ and\
  \bibinfo {author} {\bibfnamefont {Y.-C.}\ \bibnamefont {He}},\ }\bibfield
  {title} {\bibinfo {title} {Reconstructing entanglement hamiltonian via
  entanglement eigenstates},\ }\href
  {https://doi.org/10.1103/PhysRevB.99.235109} {\bibfield  {journal} {\bibinfo
  {journal} {Phys. Rev. B}\ }\textbf {\bibinfo {volume} {99}},\ \bibinfo
  {pages} {235109} (\bibinfo {year} {2019})}\BibitemShut {NoStop}%
\bibitem [{\citenamefont {Tang}\ and\ \citenamefont
  {Zhu}(2020)}]{Tang2020critical}%
  \BibitemOpen
  \bibfield  {author} {\bibinfo {author} {\bibfnamefont {Q.-C.}\ \bibnamefont
  {Tang}}\ and\ \bibinfo {author} {\bibfnamefont {W.}~\bibnamefont {Zhu}},\
  }\bibfield  {title} {\bibinfo {title} {Critical scaling behaviors of
  entanglement spectra},\ }\href
  {https://doi.org/10.1088/0256-307x/37/1/010301} {\bibfield  {journal}
  {\bibinfo  {journal} {Chinese Physics Letters}\ }\textbf {\bibinfo {volume}
  {37}},\ \bibinfo {pages} {010301} (\bibinfo {year} {2020})}\BibitemShut
  {NoStop}%
\bibitem [{\citenamefont {Mendes-Santos}\ \emph {et~al.}(2020)\citenamefont
  {Mendes-Santos}, \citenamefont {Giudici}, \citenamefont {Fazio},\ and\
  \citenamefont {Dalmonte}}]{Mendes_Santos_2020}%
  \BibitemOpen
  \bibfield  {author} {\bibinfo {author} {\bibfnamefont {T.}~\bibnamefont
  {Mendes-Santos}}, \bibinfo {author} {\bibfnamefont {G.}~\bibnamefont
  {Giudici}}, \bibinfo {author} {\bibfnamefont {R.}~\bibnamefont {Fazio}},\
  and\ \bibinfo {author} {\bibfnamefont {M.}~\bibnamefont {Dalmonte}},\
  }\bibfield  {title} {\bibinfo {title} {Measuring von neumann entanglement
  entropies without wave functions},\ }\href
  {https://doi.org/10.1088/1367-2630/ab6875} {\bibfield  {journal} {\bibinfo
  {journal} {New Journal of Physics}\ }\textbf {\bibinfo {volume} {22}},\
  \bibinfo {pages} {013044} (\bibinfo {year} {2020})}\BibitemShut {NoStop}%
\bibitem [{\citenamefont {Dalmonte}\ \emph {et~al.}(2022)\citenamefont
  {Dalmonte}, \citenamefont {Eisler}, \citenamefont {Falconi},\ and\
  \citenamefont {Vermersch}}]{Dalmonte_2022}%
  \BibitemOpen
  \bibfield  {author} {\bibinfo {author} {\bibfnamefont {M.}~\bibnamefont
  {Dalmonte}}, \bibinfo {author} {\bibfnamefont {V.}~\bibnamefont {Eisler}},
  \bibinfo {author} {\bibfnamefont {M.}~\bibnamefont {Falconi}},\ and\ \bibinfo
  {author} {\bibfnamefont {B.}~\bibnamefont {Vermersch}},\ }\bibfield  {title}
  {\bibinfo {title} {Entanglement hamiltonians: From field theory to lattice
  models and experiments},\ }\bibfield  {journal} {\bibinfo  {journal} {Annalen
  der Physik}\ }\textbf {\bibinfo {volume} {534}},\ \href
  {https://doi.org/10.1002/andp.202200064} {10.1002/andp.202200064} (\bibinfo
  {year} {2022})\BibitemShut {NoStop}%
\bibitem [{\citenamefont {Joshi}\ \emph {et~al.}(2023)\citenamefont {Joshi},
  \citenamefont {Kokail}, \citenamefont {van Bijnen}, \citenamefont {Kranzl},
  \citenamefont {Zache}, \citenamefont {Blatt}, \citenamefont {Roos},\ and\
  \citenamefont {Zoller}}]{Joshi2023}%
  \BibitemOpen
  \bibfield  {author} {\bibinfo {author} {\bibfnamefont {M.~K.}\ \bibnamefont
  {Joshi}}, \bibinfo {author} {\bibfnamefont {C.}~\bibnamefont {Kokail}},
  \bibinfo {author} {\bibfnamefont {R.}~\bibnamefont {van Bijnen}}, \bibinfo
  {author} {\bibfnamefont {F.}~\bibnamefont {Kranzl}}, \bibinfo {author}
  {\bibfnamefont {T.~V.}\ \bibnamefont {Zache}}, \bibinfo {author}
  {\bibfnamefont {R.}~\bibnamefont {Blatt}}, \bibinfo {author} {\bibfnamefont
  {C.~F.}\ \bibnamefont {Roos}},\ and\ \bibinfo {author} {\bibfnamefont
  {P.}~\bibnamefont {Zoller}},\ }\bibfield  {title} {\bibinfo {title}
  {Exploring large-scale entanglement in quantum simulation},\ }\href
  {https://doi.org/10.1038/s41586-023-06768-0} {\bibfield  {journal} {\bibinfo
  {journal} {Nature}\ }\textbf {\bibinfo {volume} {624}},\ \bibinfo {pages}
  {539} (\bibinfo {year} {2023})}\BibitemShut {NoStop}%
\bibitem [{\citenamefont {Redon}\ \emph {et~al.}(2023)\citenamefont {Redon},
  \citenamefont {Liu}, \citenamefont {Bouhiron}, \citenamefont {Mittal},
  \citenamefont {Fabre}, \citenamefont {Lopes},\ and\ \citenamefont
  {Nascimbene}}]{redon2023realizing}%
  \BibitemOpen
  \bibfield  {author} {\bibinfo {author} {\bibfnamefont {Q.}~\bibnamefont
  {Redon}}, \bibinfo {author} {\bibfnamefont {Q.}~\bibnamefont {Liu}}, \bibinfo
  {author} {\bibfnamefont {J.-B.}\ \bibnamefont {Bouhiron}}, \bibinfo {author}
  {\bibfnamefont {N.}~\bibnamefont {Mittal}}, \bibinfo {author} {\bibfnamefont
  {A.}~\bibnamefont {Fabre}}, \bibinfo {author} {\bibfnamefont
  {R.}~\bibnamefont {Lopes}},\ and\ \bibinfo {author} {\bibfnamefont
  {S.}~\bibnamefont {Nascimbene}},\ }\href@noop {} {\bibinfo {title} {Realizing
  the entanglement hamiltonian of a topological quantum hall system}} (\bibinfo
  {year} {2023}),\ \Eprint {https://arxiv.org/abs/2307.06251} {arXiv:2307.06251
  [cond-mat.quant-gas]} \BibitemShut {NoStop}%
\bibitem [{\citenamefont {Dalmonte}\ \emph {et~al.}(2018)\citenamefont
  {Dalmonte}, \citenamefont {Vermersch},\ and\ \citenamefont
  {Zoller}}]{dalmonte2018quantum}%
  \BibitemOpen
  \bibfield  {author} {\bibinfo {author} {\bibfnamefont {M.}~\bibnamefont
  {Dalmonte}}, \bibinfo {author} {\bibfnamefont {B.}~\bibnamefont
  {Vermersch}},\ and\ \bibinfo {author} {\bibfnamefont {P.}~\bibnamefont
  {Zoller}},\ }\bibfield  {title} {\bibinfo {title} {Quantum simulation and
  spectroscopy of entanglement hamiltonians},\ }\href
  {https://doi.org/10.1038/s41567-018-0151-7} {\bibfield  {journal} {\bibinfo
  {journal} {Nature Physics}\ }\textbf {\bibinfo {volume} {14}},\ \bibinfo
  {pages} {827} (\bibinfo {year} {2018})}\BibitemShut {NoStop}%
\bibitem [{\citenamefont {Ma}\ \emph {et~al.}(2023)\citenamefont {Ma},
  \citenamefont {Li},\ and\ \citenamefont {Shang}}]{ma2023multipartite}%
  \BibitemOpen
  \bibfield  {author} {\bibinfo {author} {\bibfnamefont {M.}~\bibnamefont
  {Ma}}, \bibinfo {author} {\bibfnamefont {Y.}~\bibnamefont {Li}},\ and\
  \bibinfo {author} {\bibfnamefont {J.}~\bibnamefont {Shang}},\ }\href@noop {}
  {\bibinfo {title} {Multipartite entanglement measures: a review}} (\bibinfo
  {year} {2023}),\ \Eprint {https://arxiv.org/abs/2309.09459} {arXiv:2309.09459
  [quant-ph]} \BibitemShut {NoStop}%
\bibitem [{\citenamefont {Eisler}\ \emph {et~al.}(2020)\citenamefont {Eisler},
  \citenamefont {Giulio}, \citenamefont {Tonni},\ and\ \citenamefont
  {Peschel}}]{Eisler_2020}%
  \BibitemOpen
  \bibfield  {author} {\bibinfo {author} {\bibfnamefont {V.}~\bibnamefont
  {Eisler}}, \bibinfo {author} {\bibfnamefont {G.~D.}\ \bibnamefont {Giulio}},
  \bibinfo {author} {\bibfnamefont {E.}~\bibnamefont {Tonni}},\ and\ \bibinfo
  {author} {\bibfnamefont {I.}~\bibnamefont {Peschel}},\ }\bibfield  {title}
  {\bibinfo {title} {Entanglement hamiltonians for non-critical quantum
  chains},\ }\href {https://doi.org/10.1088/1742-5468/abb4da} {\bibfield
  {journal} {\bibinfo  {journal} {Journal of Statistical Mechanics: Theory and
  Experiment}\ }\textbf {\bibinfo {volume} {2020}},\ \bibinfo {pages} {103102}
  (\bibinfo {year} {2020})}\BibitemShut {NoStop}%
\bibitem [{\citenamefont {Pouranvari}\ and\ \citenamefont
  {Abouie}(2019)}]{PhysRevB.100.195109}%
  \BibitemOpen
  \bibfield  {author} {\bibinfo {author} {\bibfnamefont {M.}~\bibnamefont
  {Pouranvari}}\ and\ \bibinfo {author} {\bibfnamefont {J.}~\bibnamefont
  {Abouie}},\ }\bibfield  {title} {\bibinfo {title} {Entanglement conductance
  as a characterization of a delocalized-localized phase transition in free
  fermion models},\ }\href {https://doi.org/10.1103/PhysRevB.100.195109}
  {\bibfield  {journal} {\bibinfo  {journal} {Phys. Rev. B}\ }\textbf {\bibinfo
  {volume} {100}},\ \bibinfo {pages} {195109} (\bibinfo {year}
  {2019})}\BibitemShut {NoStop}%
\bibitem [{\citenamefont {Cardy}\ and\ \citenamefont
  {Tonni}(2016)}]{Cardy_2016}%
  \BibitemOpen
  \bibfield  {author} {\bibinfo {author} {\bibfnamefont {J.}~\bibnamefont
  {Cardy}}\ and\ \bibinfo {author} {\bibfnamefont {E.}~\bibnamefont {Tonni}},\
  }\bibfield  {title} {\bibinfo {title} {Entanglement hamiltonians in
  two-dimensional conformal field theory},\ }\href
  {https://doi.org/10.1088/1742-5468/2016/12/123103} {\bibfield  {journal}
  {\bibinfo  {journal} {Journal of Statistical Mechanics: Theory and
  Experiment}\ }\textbf {\bibinfo {volume} {2016}},\ \bibinfo {pages} {123103}
  (\bibinfo {year} {2016})}\BibitemShut {NoStop}%
\bibitem [{\citenamefont {Javerzat}\ and\ \citenamefont
  {Tonni}(2022)}]{Javerzat_2022}%
  \BibitemOpen
  \bibfield  {author} {\bibinfo {author} {\bibfnamefont {N.}~\bibnamefont
  {Javerzat}}\ and\ \bibinfo {author} {\bibfnamefont {E.}~\bibnamefont
  {Tonni}},\ }\bibfield  {title} {\bibinfo {title} {On the continuum limit of
  the entanglement hamiltonian of a sphere for the free massless scalar
  field},\ }\bibfield  {journal} {\bibinfo  {journal} {Journal of High Energy
  Physics}\ }\textbf {\bibinfo {volume} {2022}},\ \href
  {https://doi.org/10.1007/jhep02(2022)086} {10.1007/jhep02(2022)086} (\bibinfo
  {year} {2022})\BibitemShut {NoStop}%
\bibitem [{\citenamefont {Eisler}\ \emph {et~al.}(2019)\citenamefont {Eisler},
  \citenamefont {Tonni},\ and\ \citenamefont {Peschel}}]{Eisler_2019}%
  \BibitemOpen
  \bibfield  {author} {\bibinfo {author} {\bibfnamefont {V.}~\bibnamefont
  {Eisler}}, \bibinfo {author} {\bibfnamefont {E.}~\bibnamefont {Tonni}},\ and\
  \bibinfo {author} {\bibfnamefont {I.}~\bibnamefont {Peschel}},\ }\bibfield
  {title} {\bibinfo {title} {On the continuum limit of the entanglement
  hamiltonian},\ }\href {https://doi.org/10.1088/1742-5468/ab1f0e} {\bibfield
  {journal} {\bibinfo  {journal} {Journal of Statistical Mechanics: Theory and
  Experiment}\ }\textbf {\bibinfo {volume} {2019}},\ \bibinfo {pages} {073101}
  (\bibinfo {year} {2019})}\BibitemShut {NoStop}%
\bibitem [{\citenamefont {Rao}\ \emph {et~al.}(2014)\citenamefont {Rao},
  \citenamefont {Wan},\ and\ \citenamefont {Zhang}}]{Critical}%
  \BibitemOpen
  \bibfield  {author} {\bibinfo {author} {\bibfnamefont {W.-J.}\ \bibnamefont
  {Rao}}, \bibinfo {author} {\bibfnamefont {X.}~\bibnamefont {Wan}},\ and\
  \bibinfo {author} {\bibfnamefont {G.-M.}\ \bibnamefont {Zhang}},\ }\bibfield
  {title} {\bibinfo {title} {Critical-entanglement spectrum of one-dimensional
  symmetry-protected topological phases},\ }\href
  {https://doi.org/10.1103/PhysRevB.90.075151} {\bibfield  {journal} {\bibinfo
  {journal} {Phys. Rev. B}\ }\textbf {\bibinfo {volume} {90}},\ \bibinfo
  {pages} {075151} (\bibinfo {year} {2014})}\BibitemShut {NoStop}%
\bibitem [{\citenamefont {Cirac}\ \emph
  {et~al.}(2011{\natexlab{b}})\citenamefont {Cirac}, \citenamefont {Poilblanc},
  \citenamefont {Schuch},\ and\ \citenamefont {Verstraete}}]{Entanglement2011}%
  \BibitemOpen
  \bibfield  {author} {\bibinfo {author} {\bibfnamefont {J.~I.}\ \bibnamefont
  {Cirac}}, \bibinfo {author} {\bibfnamefont {D.}~\bibnamefont {Poilblanc}},
  \bibinfo {author} {\bibfnamefont {N.}~\bibnamefont {Schuch}},\ and\ \bibinfo
  {author} {\bibfnamefont {F.}~\bibnamefont {Verstraete}},\ }\bibfield  {title}
  {\bibinfo {title} {Entanglement spectrum and boundary theories with projected
  entangled-pair states},\ }\href {https://doi.org/10.1103/PhysRevB.83.245134}
  {\bibfield  {journal} {\bibinfo  {journal} {Phys. Rev. B}\ }\textbf {\bibinfo
  {volume} {83}},\ \bibinfo {pages} {245134} (\bibinfo {year}
  {2011}{\natexlab{b}})}\BibitemShut {NoStop}%
\bibitem [{\citenamefont {He}\ \emph {et~al.}(2014)\citenamefont {He},
  \citenamefont {Sheng},\ and\ \citenamefont {Chen}}]{Chiral}%
  \BibitemOpen
  \bibfield  {author} {\bibinfo {author} {\bibfnamefont {Y.-C.}\ \bibnamefont
  {He}}, \bibinfo {author} {\bibfnamefont {D.~N.}\ \bibnamefont {Sheng}},\ and\
  \bibinfo {author} {\bibfnamefont {Y.}~\bibnamefont {Chen}},\ }\bibfield
  {title} {\bibinfo {title} {Chiral spin liquid in a frustrated anisotropic
  kagome heisenberg model},\ }\href
  {https://doi.org/10.1103/PhysRevLett.112.137202} {\bibfield  {journal}
  {\bibinfo  {journal} {Phys. Rev. Lett.}\ }\textbf {\bibinfo {volume} {112}},\
  \bibinfo {pages} {137202} (\bibinfo {year} {2014})}\BibitemShut {NoStop}%
\bibitem [{\citenamefont {Fang}\ \emph {et~al.}(2013)\citenamefont {Fang},
  \citenamefont {Gilbert},\ and\ \citenamefont {Bernevig}}]{classification}%
  \BibitemOpen
  \bibfield  {author} {\bibinfo {author} {\bibfnamefont {C.}~\bibnamefont
  {Fang}}, \bibinfo {author} {\bibfnamefont {M.~J.}\ \bibnamefont {Gilbert}},\
  and\ \bibinfo {author} {\bibfnamefont {B.~A.}\ \bibnamefont {Bernevig}},\
  }\bibfield  {title} {\bibinfo {title} {Entanglement spectrum classification
  of ${C}_{n}$-invariant noninteracting topological insulators in two
  dimensions},\ }\href {https://doi.org/10.1103/PhysRevB.87.035119} {\bibfield
  {journal} {\bibinfo  {journal} {Phys. Rev. B}\ }\textbf {\bibinfo {volume}
  {87}},\ \bibinfo {pages} {035119} (\bibinfo {year} {2013})}\BibitemShut
  {NoStop}%
\bibitem [{\citenamefont {L\"auchli}\ and\ \citenamefont
  {Schliemann}(2012)}]{lauchli2012entanglement}%
  \BibitemOpen
  \bibfield  {author} {\bibinfo {author} {\bibfnamefont {A.~M.}\ \bibnamefont
  {L\"auchli}}\ and\ \bibinfo {author} {\bibfnamefont {J.}~\bibnamefont
  {Schliemann}},\ }\bibfield  {title} {\bibinfo {title} {Entanglement spectra
  of coupled $s=\frac{1}{2}$ spin chains in a ladder geometry},\ }\href
  {https://doi.org/10.1103/PhysRevB.85.054403} {\bibfield  {journal} {\bibinfo
  {journal} {Phys. Rev. B}\ }\textbf {\bibinfo {volume} {85}},\ \bibinfo
  {pages} {054403} (\bibinfo {year} {2012})}\BibitemShut {NoStop}%
\bibitem [{\citenamefont {Diessel}\ \emph {et~al.}(2023)\citenamefont
  {Diessel}, \citenamefont {Diehl}, \citenamefont {Defenu}, \citenamefont
  {Rosch},\ and\ \citenamefont {Chiocchetta}}]{PhysRevResearch.5.033038}%
  \BibitemOpen
  \bibfield  {author} {\bibinfo {author} {\bibfnamefont {O.~K.}\ \bibnamefont
  {Diessel}}, \bibinfo {author} {\bibfnamefont {S.}~\bibnamefont {Diehl}},
  \bibinfo {author} {\bibfnamefont {N.}~\bibnamefont {Defenu}}, \bibinfo
  {author} {\bibfnamefont {A.}~\bibnamefont {Rosch}},\ and\ \bibinfo {author}
  {\bibfnamefont {A.}~\bibnamefont {Chiocchetta}},\ }\bibfield  {title}
  {\bibinfo {title} {Generalized higgs mechanism in long-range-interacting
  quantum systems},\ }\href {https://doi.org/10.1103/PhysRevResearch.5.033038}
  {\bibfield  {journal} {\bibinfo  {journal} {Phys. Rev. Res.}\ }\textbf
  {\bibinfo {volume} {5}},\ \bibinfo {pages} {033038} (\bibinfo {year}
  {2023})}\BibitemShut {NoStop}%
\bibitem [{\citenamefont {Song}\ \emph
  {et~al.}(2023{\natexlab{a}})\citenamefont {Song}, \citenamefont {Zhao},
  \citenamefont {Zhou},\ and\ \citenamefont {Meng}}]{PhysRevResearch.5.033046}%
  \BibitemOpen
  \bibfield  {author} {\bibinfo {author} {\bibfnamefont {M.}~\bibnamefont
  {Song}}, \bibinfo {author} {\bibfnamefont {J.}~\bibnamefont {Zhao}}, \bibinfo
  {author} {\bibfnamefont {C.}~\bibnamefont {Zhou}},\ and\ \bibinfo {author}
  {\bibfnamefont {Z.~Y.}\ \bibnamefont {Meng}},\ }\bibfield  {title} {\bibinfo
  {title} {Dynamical properties of quantum many-body systems with long-range
  interactions},\ }\href {https://doi.org/10.1103/PhysRevResearch.5.033046}
  {\bibfield  {journal} {\bibinfo  {journal} {Phys. Rev. Res.}\ }\textbf
  {\bibinfo {volume} {5}},\ \bibinfo {pages} {033046} (\bibinfo {year}
  {2023}{\natexlab{a}})}\BibitemShut {NoStop}%
\bibitem [{\citenamefont {Chiocchetta}\ \emph {et~al.}(2021)\citenamefont
  {Chiocchetta}, \citenamefont {Kiese}, \citenamefont {Zelle}, \citenamefont
  {Piazza},\ and\ \citenamefont {Diehl}}]{chiocchetta2021cavity}%
  \BibitemOpen
  \bibfield  {author} {\bibinfo {author} {\bibfnamefont {A.}~\bibnamefont
  {Chiocchetta}}, \bibinfo {author} {\bibfnamefont {D.}~\bibnamefont {Kiese}},
  \bibinfo {author} {\bibfnamefont {C.~P.}\ \bibnamefont {Zelle}}, \bibinfo
  {author} {\bibfnamefont {F.}~\bibnamefont {Piazza}},\ and\ \bibinfo {author}
  {\bibfnamefont {S.}~\bibnamefont {Diehl}},\ }\bibfield  {title} {\bibinfo
  {title} {Cavity-induced quantum spin liquids},\ }\href
  {https://doi.org/10.1038/s41467-021-26076-3} {\bibfield  {journal} {\bibinfo
  {journal} {Nature Communications}\ }\textbf {\bibinfo {volume} {12}},\
  \bibinfo {pages} {5901} (\bibinfo {year} {2021})}\BibitemShut {NoStop}%
\bibitem [{\citenamefont {Yusuf}\ \emph {et~al.}(2004)\citenamefont {Yusuf},
  \citenamefont {Joshi},\ and\ \citenamefont {Yang}}]{Yusuf2004spin}%
  \BibitemOpen
  \bibfield  {author} {\bibinfo {author} {\bibfnamefont {E.}~\bibnamefont
  {Yusuf}}, \bibinfo {author} {\bibfnamefont {A.}~\bibnamefont {Joshi}},\ and\
  \bibinfo {author} {\bibfnamefont {K.}~\bibnamefont {Yang}},\ }\bibfield
  {title} {\bibinfo {title} {Spin waves in antiferromagnetic spin chains with
  long-range interactions},\ }\href
  {https://doi.org/10.1103/PhysRevB.69.144412} {\bibfield  {journal} {\bibinfo
  {journal} {Phys. Rev. B}\ }\textbf {\bibinfo {volume} {69}},\ \bibinfo
  {pages} {144412} (\bibinfo {year} {2004})}\BibitemShut {NoStop}%
\bibitem [{\citenamefont {Defenu}(2021)}]{defenu2021metastability}%
  \BibitemOpen
  \bibfield  {author} {\bibinfo {author} {\bibfnamefont {N.}~\bibnamefont
  {Defenu}},\ }\bibfield  {title} {\bibinfo {title} {Metastability and discrete
  spectrum of long-range systems},\ }\href
  {https://doi.org/10.1073/pnas.2101785118} {\bibfield  {journal} {\bibinfo
  {journal} {Proceedings of the National Academy of Sciences}\ }\textbf
  {\bibinfo {volume} {118}},\ \bibinfo {pages} {e2101785118} (\bibinfo {year}
  {2021})}\BibitemShut {NoStop}%
\bibitem [{\citenamefont {Birnkammer}\ \emph {et~al.}(2022)\citenamefont
  {Birnkammer}, \citenamefont {Bohrdt}, \citenamefont {Grusdt},\ and\
  \citenamefont {Knap}}]{Birnkammer2022Characterizing}%
  \BibitemOpen
  \bibfield  {author} {\bibinfo {author} {\bibfnamefont {S.}~\bibnamefont
  {Birnkammer}}, \bibinfo {author} {\bibfnamefont {A.}~\bibnamefont {Bohrdt}},
  \bibinfo {author} {\bibfnamefont {F.}~\bibnamefont {Grusdt}},\ and\ \bibinfo
  {author} {\bibfnamefont {M.}~\bibnamefont {Knap}},\ }\bibfield  {title}
  {\bibinfo {title} {Characterizing topological excitations of a long-range
  heisenberg model with trapped ions},\ }\href
  {https://doi.org/10.1103/PhysRevB.105.L241103} {\bibfield  {journal}
  {\bibinfo  {journal} {Phys. Rev. B}\ }\textbf {\bibinfo {volume} {105}},\
  \bibinfo {pages} {L241103} (\bibinfo {year} {2022})}\BibitemShut {NoStop}%
\bibitem [{\citenamefont {Horita}\ \emph {et~al.}(2017)\citenamefont {Horita},
  \citenamefont {Suwa},\ and\ \citenamefont {Todo}}]{Horita2017upper}%
  \BibitemOpen
  \bibfield  {author} {\bibinfo {author} {\bibfnamefont {T.}~\bibnamefont
  {Horita}}, \bibinfo {author} {\bibfnamefont {H.}~\bibnamefont {Suwa}},\ and\
  \bibinfo {author} {\bibfnamefont {S.}~\bibnamefont {Todo}},\ }\bibfield
  {title} {\bibinfo {title} {Upper and lower critical decay exponents of ising
  ferromagnets with long-range interaction},\ }\href
  {https://doi.org/10.1103/PhysRevE.95.012143} {\bibfield  {journal} {\bibinfo
  {journal} {Phys. Rev. E}\ }\textbf {\bibinfo {volume} {95}},\ \bibinfo
  {pages} {012143} (\bibinfo {year} {2017})}\BibitemShut {NoStop}%
\bibitem [{\citenamefont {Fukui}\ and\ \citenamefont
  {Todo}(2009)}]{FUKUI20092629}%
  \BibitemOpen
  \bibfield  {author} {\bibinfo {author} {\bibfnamefont {K.}~\bibnamefont
  {Fukui}}\ and\ \bibinfo {author} {\bibfnamefont {S.}~\bibnamefont {Todo}},\
  }\bibfield  {title} {\bibinfo {title} {Order-n cluster monte carlo method for
  spin systems with long-range interactions},\ }\href
  {https://doi.org/https://doi.org/10.1016/j.jcp.2008.12.022} {\bibfield
  {journal} {\bibinfo  {journal} {Journal of Computational Physics}\ }\textbf
  {\bibinfo {volume} {228}},\ \bibinfo {pages} {2629} (\bibinfo {year}
  {2009})}\BibitemShut {NoStop}%
\bibitem [{\citenamefont {Mermin}\ and\ \citenamefont
  {Wagner}(1966)}]{mermin1966absence}%
  \BibitemOpen
  \bibfield  {author} {\bibinfo {author} {\bibfnamefont {N.~D.}\ \bibnamefont
  {Mermin}}\ and\ \bibinfo {author} {\bibfnamefont {H.}~\bibnamefont
  {Wagner}},\ }\bibfield  {title} {\bibinfo {title} {Absence of ferromagnetism
  or antiferromagnetism in one- or two-dimensional isotropic heisenberg
  models},\ }\href {https://doi.org/10.1103/PhysRevLett.17.1133} {\bibfield
  {journal} {\bibinfo  {journal} {Phys. Rev. Lett.}\ }\textbf {\bibinfo
  {volume} {17}},\ \bibinfo {pages} {1133} (\bibinfo {year}
  {1966})}\BibitemShut {NoStop}%
\bibitem [{\citenamefont {Mermin}(2008)}]{mermin1967absence}%
  \BibitemOpen
  \bibfield  {author} {\bibinfo {author} {\bibfnamefont {N.~D.}\ \bibnamefont
  {Mermin}},\ }\bibfield  {title} {\bibinfo {title} {{Absence of Ordering in
  Certain Classical Systems}},\ }\href {https://doi.org/10.1063/1.1705316}
  {\bibfield  {journal} {\bibinfo  {journal} {Journal of Mathematical Physics}\
  }\textbf {\bibinfo {volume} {8}},\ \bibinfo {pages} {1061} (\bibinfo {year}
  {2008})}\BibitemShut {NoStop}%
\bibitem [{\citenamefont {Anderson}(1952)}]{Anderson1952}%
  \BibitemOpen
  \bibfield  {author} {\bibinfo {author} {\bibfnamefont {P.~W.}\ \bibnamefont
  {Anderson}},\ }\bibfield  {title} {\bibinfo {title} {An approximate quantum
  theory of the antiferromagnetic ground state},\ }\href
  {https://doi.org/10.1103/PhysRev.86.694} {\bibfield  {journal} {\bibinfo
  {journal} {Phys. Rev.}\ }\textbf {\bibinfo {volume} {86}},\ \bibinfo {pages}
  {694} (\bibinfo {year} {1952})}\BibitemShut {NoStop}%
\bibitem [{\citenamefont {Oguchi}(1960)}]{Oguchi1960}%
  \BibitemOpen
  \bibfield  {author} {\bibinfo {author} {\bibfnamefont {T.}~\bibnamefont
  {Oguchi}},\ }\bibfield  {title} {\bibinfo {title} {Theory of spin-wave
  interactions in ferro- and antiferromagnetism},\ }\href
  {https://doi.org/10.1103/PhysRev.117.117} {\bibfield  {journal} {\bibinfo
  {journal} {Phys. Rev.}\ }\textbf {\bibinfo {volume} {117}},\ \bibinfo {pages}
  {117} (\bibinfo {year} {1960})}\BibitemShut {NoStop}%
\bibitem [{\citenamefont {Shao}\ \emph {et~al.}(2017)\citenamefont {Shao},
  \citenamefont {Qin}, \citenamefont {Capponi}, \citenamefont {Chesi},
  \citenamefont {Meng},\ and\ \citenamefont {Sandvik}}]{Shao2017nearly}%
  \BibitemOpen
  \bibfield  {author} {\bibinfo {author} {\bibfnamefont {H.}~\bibnamefont
  {Shao}}, \bibinfo {author} {\bibfnamefont {Y.~Q.}\ \bibnamefont {Qin}},
  \bibinfo {author} {\bibfnamefont {S.}~\bibnamefont {Capponi}}, \bibinfo
  {author} {\bibfnamefont {S.}~\bibnamefont {Chesi}}, \bibinfo {author}
  {\bibfnamefont {Z.~Y.}\ \bibnamefont {Meng}},\ and\ \bibinfo {author}
  {\bibfnamefont {A.~W.}\ \bibnamefont {Sandvik}},\ }\bibfield  {title}
  {\bibinfo {title} {Nearly deconfined spinon excitations in the square-lattice
  spin-$1/2$ heisenberg antiferromagnet},\ }\href
  {https://doi.org/10.1103/PhysRevX.7.041072} {\bibfield  {journal} {\bibinfo
  {journal} {Phys. Rev. X}\ }\textbf {\bibinfo {volume} {7}},\ \bibinfo {pages}
  {041072} (\bibinfo {year} {2017})}\BibitemShut {NoStop}%
\bibitem [{\citenamefont {Zhou}\ \emph {et~al.}(2021)\citenamefont {Zhou},
  \citenamefont {Yan}, \citenamefont {Wu}, \citenamefont {Sun}, \citenamefont
  {Starykh},\ and\ \citenamefont {Meng}}]{Zhou2021amplitude}%
  \BibitemOpen
  \bibfield  {author} {\bibinfo {author} {\bibfnamefont {C.}~\bibnamefont
  {Zhou}}, \bibinfo {author} {\bibfnamefont {Z.}~\bibnamefont {Yan}}, \bibinfo
  {author} {\bibfnamefont {H.-Q.}\ \bibnamefont {Wu}}, \bibinfo {author}
  {\bibfnamefont {K.}~\bibnamefont {Sun}}, \bibinfo {author} {\bibfnamefont
  {O.~A.}\ \bibnamefont {Starykh}},\ and\ \bibinfo {author} {\bibfnamefont
  {Z.~Y.}\ \bibnamefont {Meng}},\ }\bibfield  {title} {\bibinfo {title}
  {Amplitude mode in quantum magnets via dimensional crossover},\ }\href
  {https://doi.org/10.1103/PhysRevLett.126.227201} {\bibfield  {journal}
  {\bibinfo  {journal} {Phys. Rev. Lett.}\ }\textbf {\bibinfo {volume} {126}},\
  \bibinfo {pages} {227201} (\bibinfo {year} {2021})}\BibitemShut {NoStop}%
\bibitem [{\citenamefont {Liu}\ \emph {et~al.}(2022)\citenamefont {Liu},
  \citenamefont {Li}, \citenamefont {Huang}, \citenamefont {Li}, \citenamefont
  {Yan},\ and\ \citenamefont {Yao}}]{PhysRevB.105.014418}%
  \BibitemOpen
  \bibfield  {author} {\bibinfo {author} {\bibfnamefont {Z.}~\bibnamefont
  {Liu}}, \bibinfo {author} {\bibfnamefont {J.}~\bibnamefont {Li}}, \bibinfo
  {author} {\bibfnamefont {R.-Z.}\ \bibnamefont {Huang}}, \bibinfo {author}
  {\bibfnamefont {J.}~\bibnamefont {Li}}, \bibinfo {author} {\bibfnamefont
  {Z.}~\bibnamefont {Yan}},\ and\ \bibinfo {author} {\bibfnamefont {D.-X.}\
  \bibnamefont {Yao}},\ }\bibfield  {title} {\bibinfo {title} {Bulk and edge
  dynamics of a two-dimensional affleck-kennedy-lieb-tasaki model},\ }\href
  {https://doi.org/10.1103/PhysRevB.105.014418} {\bibfield  {journal} {\bibinfo
   {journal} {Phys. Rev. B}\ }\textbf {\bibinfo {volume} {105}},\ \bibinfo
  {pages} {014418} (\bibinfo {year} {2022})}\BibitemShut {NoStop}%
\bibitem [{\citenamefont {Chandran}\ \emph {et~al.}(2014)\citenamefont
  {Chandran}, \citenamefont {Khemani},\ and\ \citenamefont
  {Sondhi}}]{Chandran2014how}%
  \BibitemOpen
  \bibfield  {author} {\bibinfo {author} {\bibfnamefont {A.}~\bibnamefont
  {Chandran}}, \bibinfo {author} {\bibfnamefont {V.}~\bibnamefont {Khemani}},\
  and\ \bibinfo {author} {\bibfnamefont {S.~L.}\ \bibnamefont {Sondhi}},\
  }\bibfield  {title} {\bibinfo {title} {How universal is the entanglement
  spectrum?},\ }\href {https://doi.org/10.1103/PhysRevLett.113.060501}
  {\bibfield  {journal} {\bibinfo  {journal} {Phys. Rev. Lett.}\ }\textbf
  {\bibinfo {volume} {113}},\ \bibinfo {pages} {060501} (\bibinfo {year}
  {2014})}\BibitemShut {NoStop}%
\bibitem [{\citenamefont {Yan}\ and\ \citenamefont
  {Meng}(2023)}]{zyan2021entanglement}%
  \BibitemOpen
  \bibfield  {author} {\bibinfo {author} {\bibfnamefont {Z.}~\bibnamefont
  {Yan}}\ and\ \bibinfo {author} {\bibfnamefont {Z.~Y.}\ \bibnamefont {Meng}},\
  }\bibfield  {title} {\bibinfo {title} {Unlocking the general relationship
  between energy and entanglement spectra via the wormhole effect},\ }\href
  {https://doi.org/10.1038/s41467-023-37756-7} {\bibfield  {journal} {\bibinfo
  {journal} {Nature Communications}\ }\textbf {\bibinfo {volume} {14}},\
  \bibinfo {pages} {2360} (\bibinfo {year} {2023})}\BibitemShut {NoStop}%
\bibitem [{\citenamefont {Zhao}\ \emph {et~al.}(2023)\citenamefont {Zhao},
  \citenamefont {Song}, \citenamefont {Qi}, \citenamefont {Rong},\ and\
  \citenamefont {Meng}}]{Zhao2023}%
  \BibitemOpen
  \bibfield  {author} {\bibinfo {author} {\bibfnamefont {J.}~\bibnamefont
  {Zhao}}, \bibinfo {author} {\bibfnamefont {M.}~\bibnamefont {Song}}, \bibinfo
  {author} {\bibfnamefont {Y.}~\bibnamefont {Qi}}, \bibinfo {author}
  {\bibfnamefont {J.}~\bibnamefont {Rong}},\ and\ \bibinfo {author}
  {\bibfnamefont {Z.~Y.}\ \bibnamefont {Meng}},\ }\bibfield  {title} {\bibinfo
  {title} {Finite-temperature critical behaviors in 2d long-range quantum
  heisenberg model},\ }\href {https://doi.org/10.1038/s41535-023-00591-6}
  {\bibfield  {journal} {\bibinfo  {journal} {npj Quantum Materials}\ }\textbf
  {\bibinfo {volume} {8}},\ \bibinfo {pages} {59} (\bibinfo {year}
  {2023})}\BibitemShut {NoStop}%
\bibitem [{\citenamefont {Song}\ \emph
  {et~al.}(2023{\natexlab{b}})\citenamefont {Song}, \citenamefont {Zhao},
  \citenamefont {Qi}, \citenamefont {Rong},\ and\ \citenamefont
  {Meng}}]{song2023quantum}%
  \BibitemOpen
  \bibfield  {author} {\bibinfo {author} {\bibfnamefont {M.}~\bibnamefont
  {Song}}, \bibinfo {author} {\bibfnamefont {J.}~\bibnamefont {Zhao}}, \bibinfo
  {author} {\bibfnamefont {Y.}~\bibnamefont {Qi}}, \bibinfo {author}
  {\bibfnamefont {J.}~\bibnamefont {Rong}},\ and\ \bibinfo {author}
  {\bibfnamefont {Z.~Y.}\ \bibnamefont {Meng}},\ }\href@noop {} {\bibinfo
  {title} {Quantum criticality and entanglement for 2d long-range heisenberg
  bilayer}} (\bibinfo {year} {2023}{\natexlab{b}}),\ \Eprint
  {https://arxiv.org/abs/2306.05465} {arXiv:2306.05465 [cond-mat.str-el]}
  \BibitemShut {NoStop}%
\bibitem [{\citenamefont
  {Sandvik}(1998{\natexlab{a}})}]{sandvik1998stochastic}%
  \BibitemOpen
  \bibfield  {author} {\bibinfo {author} {\bibfnamefont {A.~W.}\ \bibnamefont
  {Sandvik}},\ }\bibfield  {title} {\bibinfo {title} {Stochastic method for
  analytic continuation of quantum monte carlo data},\ }\href
  {https://doi.org/10.1103/PhysRevB.57.10287} {\bibfield  {journal} {\bibinfo
  {journal} {Phys. Rev. B}\ }\textbf {\bibinfo {volume} {57}},\ \bibinfo
  {pages} {10287} (\bibinfo {year} {1998}{\natexlab{a}})}\BibitemShut {NoStop}%
\bibitem [{\citenamefont {Sandvik}(2010)}]{sandvik2010computational}%
  \BibitemOpen
  \bibfield  {author} {\bibinfo {author} {\bibfnamefont {A.~W.}\ \bibnamefont
  {Sandvik}},\ }\bibfield  {title} {\bibinfo {title} {{Computational Studies of
  Quantum Spin Systems}},\ }\href {https://doi.org/10.1063/1.3518900}
  {\bibfield  {journal} {\bibinfo  {journal} {AIP Conference Proceedings}\
  }\textbf {\bibinfo {volume} {1297}},\ \bibinfo {pages} {135} (\bibinfo {year}
  {2010})}\BibitemShut {NoStop}%
\bibitem [{\citenamefont {Sandvik}\ and\ \citenamefont
  {Kurkij\"arvi}(1991)}]{Sandvik1991}%
  \BibitemOpen
  \bibfield  {author} {\bibinfo {author} {\bibfnamefont {A.~W.}\ \bibnamefont
  {Sandvik}}\ and\ \bibinfo {author} {\bibfnamefont {J.}~\bibnamefont
  {Kurkij\"arvi}},\ }\bibfield  {title} {\bibinfo {title} {Quantum {M}onte
  {C}arlo simulation method for spin systems},\ }\href
  {https://doi.org/10.1103/PhysRevB.43.5950} {\bibfield  {journal} {\bibinfo
  {journal} {Phys. Rev. B}\ }\textbf {\bibinfo {volume} {43}},\ \bibinfo
  {pages} {5950} (\bibinfo {year} {1991})}\BibitemShut {NoStop}%
\bibitem [{\citenamefont {Sandvik}(1999)}]{Sandvik1999}%
  \BibitemOpen
  \bibfield  {author} {\bibinfo {author} {\bibfnamefont {A.~W.}\ \bibnamefont
  {Sandvik}},\ }\bibfield  {title} {\bibinfo {title} {Stochastic series
  expansion method with operator-loop update},\ }\href
  {https://doi.org/10.1103/PhysRevB.59.R14157} {\bibfield  {journal} {\bibinfo
  {journal} {Phys. Rev. B}\ }\textbf {\bibinfo {volume} {59}},\ \bibinfo
  {pages} {R14157} (\bibinfo {year} {1999})}\BibitemShut {NoStop}%
\bibitem [{\citenamefont {Sylju\aa{}sen}\ and\ \citenamefont
  {Sandvik}(2002)}]{Syljuaasen2002}%
  \BibitemOpen
  \bibfield  {author} {\bibinfo {author} {\bibfnamefont {O.~F.}\ \bibnamefont
  {Sylju\aa{}sen}}\ and\ \bibinfo {author} {\bibfnamefont {A.~W.}\ \bibnamefont
  {Sandvik}},\ }\bibfield  {title} {\bibinfo {title} {Quantum monte carlo with
  directed loops},\ }\href {https://doi.org/10.1103/PhysRevE.66.046701}
  {\bibfield  {journal} {\bibinfo  {journal} {Phys. Rev. E}\ }\textbf {\bibinfo
  {volume} {66}},\ \bibinfo {pages} {046701} (\bibinfo {year}
  {2002})}\BibitemShut {NoStop}%
\bibitem [{\citenamefont {Yan}\ \emph {et~al.}(2019)\citenamefont {Yan},
  \citenamefont {Wu}, \citenamefont {Liu}, \citenamefont {Sylju\aa{}sen},
  \citenamefont {Lou},\ and\ \citenamefont {Chen}}]{ZY2019}%
  \BibitemOpen
  \bibfield  {author} {\bibinfo {author} {\bibfnamefont {Z.}~\bibnamefont
  {Yan}}, \bibinfo {author} {\bibfnamefont {Y.}~\bibnamefont {Wu}}, \bibinfo
  {author} {\bibfnamefont {C.}~\bibnamefont {Liu}}, \bibinfo {author}
  {\bibfnamefont {O.~F.}\ \bibnamefont {Sylju\aa{}sen}}, \bibinfo {author}
  {\bibfnamefont {J.}~\bibnamefont {Lou}},\ and\ \bibinfo {author}
  {\bibfnamefont {Y.}~\bibnamefont {Chen}},\ }\bibfield  {title} {\bibinfo
  {title} {Sweeping cluster algorithm for quantum spin systems with strong
  geometric restrictions},\ }\href {https://doi.org/10.1103/PhysRevB.99.165135}
  {\bibfield  {journal} {\bibinfo  {journal} {Phys. Rev. B}\ }\textbf {\bibinfo
  {volume} {99}},\ \bibinfo {pages} {165135} (\bibinfo {year}
  {2019})}\BibitemShut {NoStop}%
\bibitem [{\citenamefont {Yan}(2022)}]{yan2020improved}%
  \BibitemOpen
  \bibfield  {author} {\bibinfo {author} {\bibfnamefont {Z.}~\bibnamefont
  {Yan}},\ }\bibfield  {title} {\bibinfo {title} {Global scheme of sweeping
  cluster algorithm to sample among topological sectors},\ }\href
  {https://doi.org/10.1103/PhysRevB.105.184432} {\bibfield  {journal} {\bibinfo
   {journal} {Phys. Rev. B}\ }\textbf {\bibinfo {volume} {105}},\ \bibinfo
  {pages} {184432} (\bibinfo {year} {2022})}\BibitemShut {NoStop}%
\bibitem [{\citenamefont {Liu}\ \emph {et~al.}(2023)\citenamefont {Liu},
  \citenamefont {Huang}, \citenamefont {Yan},\ and\ \citenamefont
  {Yao}}]{liu2023probing}%
  \BibitemOpen
  \bibfield  {author} {\bibinfo {author} {\bibfnamefont {Z.}~\bibnamefont
  {Liu}}, \bibinfo {author} {\bibfnamefont {R.-Z.}\ \bibnamefont {Huang}},
  \bibinfo {author} {\bibfnamefont {Z.}~\bibnamefont {Yan}},\ and\ \bibinfo
  {author} {\bibfnamefont {D.-X.}\ \bibnamefont {Yao}},\ }\bibfield  {title}
  {\bibinfo {title} {Probing li-haldane conjecture with a perturbed boundary},\
  }\href@noop {} {\  (\bibinfo {year} {2023})},\ \Eprint
  {https://arxiv.org/abs/2303.00772} {arXiv:2303.00772 [cond-mat.str-el]}
  \BibitemShut {NoStop}%
\bibitem [{\citenamefont {Wu}\ \emph {et~al.}(2023)\citenamefont {Wu},
  \citenamefont {Ran}, \citenamefont {Yin}, \citenamefont {Li}, \citenamefont
  {Mao}, \citenamefont {Wang},\ and\ \citenamefont {Yan}}]{wu2023classical}%
  \BibitemOpen
  \bibfield  {author} {\bibinfo {author} {\bibfnamefont {S.}~\bibnamefont
  {Wu}}, \bibinfo {author} {\bibfnamefont {X.}~\bibnamefont {Ran}}, \bibinfo
  {author} {\bibfnamefont {B.}~\bibnamefont {Yin}}, \bibinfo {author}
  {\bibfnamefont {Q.-F.}\ \bibnamefont {Li}}, \bibinfo {author} {\bibfnamefont
  {B.-B.}\ \bibnamefont {Mao}}, \bibinfo {author} {\bibfnamefont {Y.-C.}\
  \bibnamefont {Wang}},\ and\ \bibinfo {author} {\bibfnamefont
  {Z.}~\bibnamefont {Yan}},\ }\bibfield  {title} {\bibinfo {title} {Classical
  model emerges in quantum entanglement: Quantum monte carlo study for an
  ising-heisenberg bilayer},\ }\href
  {https://doi.org/10.1103/PhysRevB.107.155121} {\bibfield  {journal} {\bibinfo
   {journal} {Phys. Rev. B}\ }\textbf {\bibinfo {volume} {107}},\ \bibinfo
  {pages} {155121} (\bibinfo {year} {2023})}\BibitemShut {NoStop}%
\bibitem [{\citenamefont {Song}\ \emph
  {et~al.}(2023{\natexlab{c}})\citenamefont {Song}, \citenamefont {Zhao},
  \citenamefont {Yan},\ and\ \citenamefont {Meng}}]{song2023different}%
  \BibitemOpen
  \bibfield  {author} {\bibinfo {author} {\bibfnamefont {M.}~\bibnamefont
  {Song}}, \bibinfo {author} {\bibfnamefont {J.}~\bibnamefont {Zhao}}, \bibinfo
  {author} {\bibfnamefont {Z.}~\bibnamefont {Yan}},\ and\ \bibinfo {author}
  {\bibfnamefont {Z.~Y.}\ \bibnamefont {Meng}},\ }\bibfield  {title} {\bibinfo
  {title} {Different temperature dependence for the edge and bulk of the
  entanglement hamiltonian},\ }\href
  {https://doi.org/10.1103/PhysRevB.108.075114} {\bibfield  {journal} {\bibinfo
   {journal} {Phys. Rev. B}\ }\textbf {\bibinfo {volume} {108}},\ \bibinfo
  {pages} {075114} (\bibinfo {year} {2023}{\natexlab{c}})}\BibitemShut
  {NoStop}%
\bibitem [{\citenamefont {Sandvik}(1998{\natexlab{b}})}]{Sandvik1998a}%
  \BibitemOpen
  \bibfield  {author} {\bibinfo {author} {\bibfnamefont {A.~W.}\ \bibnamefont
  {Sandvik}},\ }\bibfield  {title} {\bibinfo {title} {{Stochastic method for
  analytic continuation of quantum Monte Carlo data}},\ }\href
  {https://doi.org/10.1103/PhysRevB.57.10287} {\bibfield  {journal} {\bibinfo
  {journal} {Phys. Rev. B}\ }\textbf {\bibinfo {volume} {57}},\ \bibinfo
  {pages} {10287} (\bibinfo {year} {1998}{\natexlab{b}})}\BibitemShut {NoStop}%
\bibitem [{\citenamefont {{Beach}}(2004)}]{Beach2004}%
  \BibitemOpen
  \bibfield  {author} {\bibinfo {author} {\bibfnamefont {K.~S.~D.}\
  \bibnamefont {{Beach}}},\ }\bibfield  {title} {\bibinfo {title} {{Identifying
  the maximum entropy method as a special limit of stochastic analytic
  continuation}},\ }\href@noop {} {\bibfield  {journal} {\bibinfo  {journal}
  {arXiv e-prints}\ ,\ \bibinfo {eid} {cond-mat/0403055}} (\bibinfo {year}
  {2004})},\ \Eprint {https://arxiv.org/abs/cond-mat/0403055}
  {arXiv:cond-mat/0403055 [cond-mat.str-el]} \BibitemShut {NoStop}%
\bibitem [{\citenamefont {Sylju\aa{}sen}(2008)}]{Syljuasen2008}%
  \BibitemOpen
  \bibfield  {author} {\bibinfo {author} {\bibfnamefont {O.~F.}\ \bibnamefont
  {Sylju\aa{}sen}},\ }\bibfield  {title} {\bibinfo {title} {Using the average
  spectrum method to extract dynamics from quantum monte carlo simulations},\
  }\href {https://doi.org/10.1103/PhysRevB.78.174429} {\bibfield  {journal}
  {\bibinfo  {journal} {Phys. Rev. B}\ }\textbf {\bibinfo {volume} {78}},\
  \bibinfo {pages} {174429} (\bibinfo {year} {2008})}\BibitemShut {NoStop}%
\bibitem [{\citenamefont {Shao}\ and\ \citenamefont
  {Sandvik}(2023)}]{SHAO2023Progress}%
  \BibitemOpen
  \bibfield  {author} {\bibinfo {author} {\bibfnamefont {H.}~\bibnamefont
  {Shao}}\ and\ \bibinfo {author} {\bibfnamefont {A.~W.}\ \bibnamefont
  {Sandvik}},\ }\href
  {https://doi.org/https://doi.org/10.1016/j.physrep.2022.11.002} {\bibfield
  {journal} {\bibinfo  {journal} {Physics Reports}\ }\textbf {\bibinfo {volume}
  {1003}},\ \bibinfo {pages} {1} (\bibinfo {year} {2023})}\BibitemShut
  {NoStop}%
\bibitem [{\citenamefont {Yoshino}\ \emph {et~al.}(2021)\citenamefont
  {Yoshino}, \citenamefont {Furukawa},\ and\ \citenamefont
  {Ueda}}]{PhysRevA.103.043321}%
  \BibitemOpen
  \bibfield  {author} {\bibinfo {author} {\bibfnamefont {T.}~\bibnamefont
  {Yoshino}}, \bibinfo {author} {\bibfnamefont {S.}~\bibnamefont {Furukawa}},\
  and\ \bibinfo {author} {\bibfnamefont {M.}~\bibnamefont {Ueda}},\ }\bibfield
  {title} {\bibinfo {title} {Intercomponent entanglement entropy and spectrum
  in binary bose-einstein condensates},\ }\href
  {https://doi.org/10.1103/PhysRevA.103.043321} {\bibfield  {journal} {\bibinfo
   {journal} {Phys. Rev. A}\ }\textbf {\bibinfo {volume} {103}},\ \bibinfo
  {pages} {043321} (\bibinfo {year} {2021})}\BibitemShut {NoStop}%
\bibitem [{\citenamefont {Lundgren}\ \emph {et~al.}(2013)\citenamefont
  {Lundgren}, \citenamefont {Fuji}, \citenamefont {Furukawa},\ and\
  \citenamefont {Oshikawa}}]{PhysRevB.88.245137}%
  \BibitemOpen
  \bibfield  {author} {\bibinfo {author} {\bibfnamefont {R.}~\bibnamefont
  {Lundgren}}, \bibinfo {author} {\bibfnamefont {Y.}~\bibnamefont {Fuji}},
  \bibinfo {author} {\bibfnamefont {S.}~\bibnamefont {Furukawa}},\ and\
  \bibinfo {author} {\bibfnamefont {M.}~\bibnamefont {Oshikawa}},\ }\bibfield
  {title} {\bibinfo {title} {Entanglement spectra between coupled
  tomonaga-luttinger liquids: Applications to ladder systems and topological
  phases},\ }\href {https://doi.org/10.1103/PhysRevB.88.245137} {\bibfield
  {journal} {\bibinfo  {journal} {Phys. Rev. B}\ }\textbf {\bibinfo {volume}
  {88}},\ \bibinfo {pages} {245137} (\bibinfo {year} {2013})}\BibitemShut
  {NoStop}%
\bibitem [{\citenamefont {Zhou}\ \emph {et~al.}(2023)\citenamefont {Zhou},
  \citenamefont {Li}, \citenamefont {Zhai}, \citenamefont {Li},\ and\
  \citenamefont {Gu}}]{zhou2023reviving}%
  \BibitemOpen
  \bibfield  {author} {\bibinfo {author} {\bibfnamefont {Y.-N.}\ \bibnamefont
  {Zhou}}, \bibinfo {author} {\bibfnamefont {X.}~\bibnamefont {Li}}, \bibinfo
  {author} {\bibfnamefont {H.}~\bibnamefont {Zhai}}, \bibinfo {author}
  {\bibfnamefont {C.}~\bibnamefont {Li}},\ and\ \bibinfo {author}
  {\bibfnamefont {Y.}~\bibnamefont {Gu}},\ }\href@noop {} {\bibinfo {title}
  {Reviving the lieb-schultz-mattis theorem in open quantum systems}} (\bibinfo
  {year} {2023}),\ \Eprint {https://arxiv.org/abs/2310.01475} {arXiv:2310.01475
  [cond-mat.str-el]} \BibitemShut {NoStop}%
\bibitem [{\citenamefont {Li}\ \emph {et~al.}(2023)\citenamefont {Li},
  \citenamefont {Li},\ and\ \citenamefont {Zhou}}]{li2023numerical}%
  \BibitemOpen
  \bibfield  {author} {\bibinfo {author} {\bibfnamefont {C.}~\bibnamefont
  {Li}}, \bibinfo {author} {\bibfnamefont {X.}~\bibnamefont {Li}},\ and\
  \bibinfo {author} {\bibfnamefont {Y.-N.}\ \bibnamefont {Zhou}},\ }\href@noop
  {} {\bibinfo {title} {Numerical investigations of the extensive entanglement
  hamiltonian in quantum spin ladders}} (\bibinfo {year} {2023}),\ \Eprint
  {https://arxiv.org/abs/2311.01699} {arXiv:2311.01699 [cond-mat.str-el]}
  \BibitemShut {NoStop}%
\end{thebibliography}%

\end{document}